\documentclass[journal]{IEEEtran}

\usepackage{cite}
\usepackage{amsthm}
\usepackage{amsmath}
\usepackage{algorithm}
\usepackage{algpseudocode}
\usepackage{comment}
\usepackage{booktabs}
\usepackage{longtable}
\usepackage{stfloats}
\usepackage{bm}
\usepackage{setspace}
\usepackage{graphicx}
\usepackage[usenames,dvipsnames]{color}
\usepackage{amsfonts,amssymb,amsmath}
\usepackage{enumerate}
\usepackage{cases}
\usepackage{array}
\usepackage{colortbl}
\usepackage{slashbox}
\usepackage{multirow}
\usepackage{arydshln}
\usepackage{caption}
\usepackage{float}
\usepackage{url}
\usepackage{multirow}

\usepackage{graphicx}
\usepackage{subcaption}
\usepackage{wasysym}

\definecolor{Gray}{gray}{0.85}
\definecolor{mycyan}{cmyk}{.3,0,0,0}

\newcolumntype{a}{>{\columncolor{Gray}}c}
\newcolumntype{b}{>{\columncolor{white}}c}
\newcolumntype{d}{>{\columncolor{mycyan}}c}

\allowdisplaybreaks
\begin{document}
\captionsetup[figure]{name={Fig.},labelsep=period}
\title{Robust DNN Partitioning and Resource Allocation Under Uncertain Inference Time}
\author{Zhaojun~Nan,~\IEEEmembership{Member,~IEEE,}
        Yunchu~Han,~\IEEEmembership{Student Member,~IEEE,}
        Sheng~Zhou,~\IEEEmembership{Senior Member,~IEEE,}
        and~Zhisheng Niu,~\IEEEmembership{Fellow,~IEEE}

\thanks{This work is supported in part by the National Natural Science Foundation of China under Grants 62341108, in part by the China Postdoctoral Science Foundation under Grant 2023M742011, and in part by Hitachi Ltd. (\emph{Corresponding author: Sheng Zhou.)}
}

\thanks{Zhaojun Nan, Yunchu Han, Sheng Zhou, and Zhisheng Niu are with the Beijing National Research Center for Information Science and Technology, Department of Electronic Engineering, Tsinghua University, Beijing 100084, China (e-mail: nzj660624@mail.tsinghua.edu.cn; hyc23@mails.tsinghua.edu.cn; sheng.zhou@tsinghua.edu.cn; niuzhs@tsinghua\-.edu.cn).
}}

\maketitle


\vspace{-3em}
\begin{abstract}
In edge intelligence systems, deep neural network (DNN) partitioning and data offloading can provide real-time task inference for resource-constrained mobile devices. However, the inference time of DNNs is typically uncertain and cannot be precisely determined in advance, presenting significant challenges in ensuring timely task processing within deadlines. To address the uncertain inference time, we propose a robust optimization scheme to minimize the total energy consumption of mobile devices while meeting task probabilistic deadlines. The scheme only requires the mean and variance information of the inference time, without any prediction methods or distribution functions. The problem is formulated as a mixed-integer nonlinear programming (MINLP) that involves jointly optimizing the DNN model partitioning and the allocation of local CPU/GPU frequencies and uplink bandwidth. To tackle the problem, we first decompose the original problem into two subproblems: resource allocation and DNN model partitioning. Subsequently, the two subproblems with probability constraints are equivalently transformed into deterministic optimization problems using the chance-constrained programming (CCP) method. Finally, the convex optimization technique and the penalty convex-concave procedure (PCCP) technique are employed to obtain the optimal solution of the resource allocation subproblem and a stationary point of the DNN model partitioning subproblem, respectively. The proposed algorithm leverages real-world data from popular hardware platforms and is evaluated on widely used DNN models. Extensive simulations show that our proposed algorithm effectively addresses the inference time uncertainty with probabilistic deadline guarantees while minimizing the energy consumption of mobile devices.
\end{abstract}


\section{Introduction}
\IEEEPARstart {D}{eep} neural networks (DNNs) have been extensively applied across various innovative applications, including speech recognition \cite{Amodei1}, object detection \cite{Tan2}, image segmentation \cite{Chen3}, etc. With the penetration of these applications, there is a critical demand to deploy DNN models on mobile devices with limited computing capacity and battery power, such as energy-harvesting sensors, micro-robots, and unmanned aerial vehicles, to achieve real-time task inference and intelligent decision-making. However, these DNN models usually have high computing capacity requirements. For example, GoogleNet, ResNet101, and VGG16 require 3.0, 15.2, and 31.0 giga floating point of operations (GFLOPs), respectively \cite{Desislavov4}. On the Raspberry Pi platform, the inference time of GoogleNet is about $0.8$ seconds \cite{Li5}, while for tiny YOLOv2, it is up to $1.8$ seconds \cite{Hu6}. Due to the disparity between the high computing capacity requirements of DNNs and the resource-limited mobile devices, achieving fast task inference on these mobile devices is highly challenging.

To address this challenge, edge-device collaborative inference has recently been proposed \cite{Hu6}, \cite{Kang7}. The key idea of edge-device collaborative inference is to adaptively partition the DNN model in response to varying channel states, thereby achieving an efficient balance of the inference computing capacity and transmission data size between mobile devices and the edge server. This facilitates the coordination of timely task inference between weak mobile devices and the powerful edge server. The important objective of collaborative inference is to determine the optimal partitioning point and allocate communication and computation resources, ensuring that the inference results meet task deadlines and enabling the timely processing of subsequent tasks. However, most existing work on collaborative inference assumes that the inference time of a task is precisely known, overlooking the impact of inference time uncertainty on collaborative inference \cite{Tang8}, \cite{Zeng9}, \cite{Zhang10}, \cite{Mohammed11}, \cite{Shi12}, \cite{Su13}, \cite{Li14}, \cite{Xu15}.

\begin{figure}[t]
  \captionsetup{font={small}}
  \centering
  \includegraphics[width=0.50\textwidth]{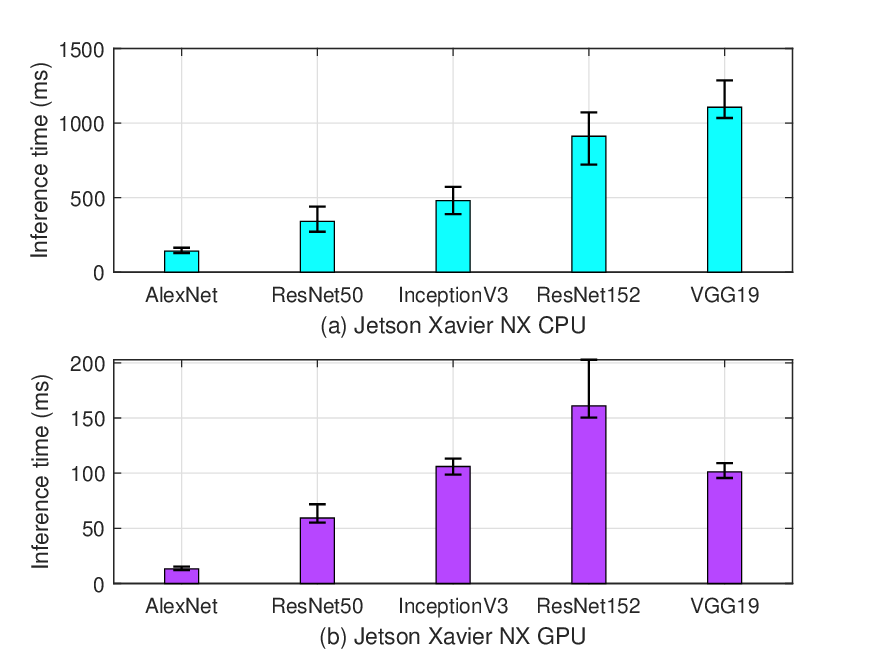}\\
  \caption{The variation in inference time for the classification task on the CIFAR-10 dataset using the CPU and GPU of the NVIDIA Jetson Xavier NX platform, respectively.}
  \label{fig:Variation}
  \vspace{-0.1cm}
\end{figure}

In practical systems, the inference time of DNNs is variable and uncertain, and it cannot be determined until the inference task is executed \cite{Wu16}, \cite{{Liu17}}. In \cite{Wu16}, the authors assess the inference time of convolutional neural networks on the SoCs, observing significant performance variations under inference time outliers. In \cite{Liu17}, the authors observe that the inference time of various DNN models applied to autonomous driving are uncertain, and analyze several factors that influence the fluctuations in DNN inference time. Different from the object detection tasks in \cite{Liu17}, we verify the variations in inference time of several DNN models for the classification task on the CIFAR-10 dataset using the CPU and GPU platforms, as shown in Fig. \ref{fig:Variation}. We also find that the uncertainty of DNN inference time is affected by the model structure, I/O speed, hardware platform, etc. Moreover, it can be observed from Fig. \ref{fig:Variation} that the inference time of different models on different hardware exhibits significant randomness, which makes its distribution function difficult to obtain accurately. Indeed, uncertain inference time brings a significant challenge to edge-device collaboration. It is well known that deciding DNN model partitioning based on worst-case inference time tends to be overly conservative, and extending task deadlines can compromise the timeliness of the system. Therefore, it is necessary to consider the robust DNN partitioning and resource allocation to provide performance guarantees.

In this paper, we address the issue of uncertain inference time in DNN model partitioning and resource allocation by providing probabilistic guarantees on deadlines. In this way, inference time is not strictly bound by hard deadlines; rather, occasional violations of task deadlines are tolerated. This approach is deemed reasonable in practical systems. For image or video processing, occasional violations of deadlines can be mitigated through error control techniques at the application layer \cite{Wu33}. Specifically, we allow the probability of task execution time (consisting of local inference, uplink transmission, and edge inference delays) violating the task deadline to remain under a predefined threshold while minimizing the total energy consumption of mobile devices. Considering that the accurate inference time cannot be obtained and its distribution function is difficult to characterize, we design a robust DNN partitioning, uplink bandwidth, and computing resource allocation policy, utilizing only the mean and variance information of the inference time. The main contributions of this work are summarized below.

\begin{itemize}
\item
To the best of our knowledge, this is the first work explicitly considering inference time uncertainty in optimizing DNN partitioning. To this end, we formulate a joint optimization problem involving the DNN model partitioning and the allocation of local CPU/GPU frequencies and uplink bandwidth under uncertain inference time, aiming to minimize the expected energy consumption of all mobile devices while meeting probabilistic deadline constraints. Due to the probabilistic deadline constraints arising from uncertain inference time and the combinatorial nature of DNN model partitioning and resource allocation decisions, the problem is a challenging mixed-integer nonlinear programming (MINLP) problem.
\item
Considering that DNN inference time cannot be precisely determined a $\emph{priori}$ and its probabilistic distribution is difficult to estimate accurately, we characterize the mean and variance of the inference time across different CPU/GPU frequencies based on real-world data from DNN models. Specifically, the nonlinear least squares method is used to fit a function that describes the relationship between the mean inference time and CPU/GPU frequency. Then, we present an efficient method for estimating the variance and covariance of the inference time across different CPU/GPU frequencies.
\item
To deal with the combinatorial nature of the MINLP problem, we first propose decomposing the original problem into a resource allocation subproblem with fixed partitioning decisions and a DNN model partitioning subproblem that optimizes the expected energy consumption corresponding to the resource allocation problem. Then, the two subproblems with probabilistic constraints are equivalently transformed into deterministic optimization problems using the mean and variance information of inference time and the chance-constrained programming (CCP) method.
\item
Finally, we obtain the optimal solution to uplink bandwidth and the CPU/GPU frequencies of the resource allocation subproblem using the convex optimization technique. By exploring the structural properties of the DNN model partitioning subproblem, a stationary point of the problem is obtained using the penalty convex-concave procedure (PCCP) method. The PCCP method has low computational complexity and can achieve the near-optimal solution in polynomial time.
\end{itemize}

Simulations are carried out on real-world data from Nvidia hardware platforms and are evaluated on widely used DNN models. Through extensive simulations, we demonstrate that the proposed robust policy exhibits faster convergence and lower computational complexity. The simulation results show that the probability guarantee of the task deadline can be successfully achieved under DNN inference time uncertainty, which means that the proposed policy is more robust. Compared to the state-of-the-art approach, our proposed policy has a significant improvement in energy saving on mobile devices.

The remainder of this paper is organized as follows. Section II reviews related work. Section III describes the system model and problem formulation. Section IV derives the mean and variance of inference time. Section V develops a robust DNN partitioning and resource allocation algorithm. Section VI shows the simulation results, followed by the conclusion in Section VII.

\section{Related Work}
In this section, we summarize the existing work on DNN model partitioning and resource allocation, and introduce the work related to inference time uncertainty.

\subsection{DNN Model Partitioning}
Extensive research focuses on DNN model partitioning and resource allocation in collaboration inference. Various works are investigated from different perspectives, such as collaborative paradigms, DNN model structures, and inference approaches. The cloud-end collaboration paradigm partially shifts DNN inference from the device to the cloud \cite{Kang7}, \cite{Teerapittayanon18}. In \cite{Kang7}, the neurosurgeon algorithm is proposed to find an intermediate partitioning point in the DNN model, keeping the front-end model on the device and offloading the back-end model to the cloud. Leveraging a similar principle, \cite{Teerapittayanon18} proposes a distributed partitioning strategy that divides the DNN model into the cloud, the edge server, and the end devices. To reduce the latency of cloud inference, the edge-end collaboration paradigm is studied \cite{Li5}, \cite{Xu15}. Edgent \cite{Li5} utilizes mobile edge computing (MEC) for DNN collaborative inference through device-edge synergy by adaptive partitioning and right-sizing DNNs to reduce latency. Based on \cite{Li5}, \cite{Xu15} proposes a learning-based method that optimizes DNN partitioning, early exit point selection, and computing resource allocation. \cite{Wen45} proposes an integrated sensing, computing, and communication (ISCC) architecture that jointly optimizes AI model partitioning, as well as integrated sensing and communication, to deliver low-latency intelligent services at the network edge.

Different DNNs may have various structures, so a suitable model structure is needed for effective partitioning. Therefore, \cite{Hu6} and \cite{Li14} use the directed acyclic graph (DAG) to model the relationship between layers in DNN, and transform the DNN partitioning problem into the solution of the minimum cut problem in graph theory. To reduce the complexity of DAG modeling, \cite{Zhang10} and \cite{Zhao19} divide the DAG into multiple blocks, thereby simplifying the DNN model into a block-based chain structure. The above studies generally adopt a sequence inference approach, where local inference is before the partitioning point and edge inference is behind the partitioning point. Unlike sequence inference, a few works investigate parallel and batch inference approaches. Taking advantage of the parallelism of the input sequence, \cite{Hu20} partitions the transformer model according to location to accelerate the inference speed. \cite{Shi12} considers appropriate partitioning point selection, aggregates multiple inference tasks into one batch, and processes them concurrently on the edge server. However, most of these studies assume that the inference time of DNNs is deterministic and known in advance.

\subsection{Inference Time Uncertainty}
A few works that focus on the inference time uncertainty \cite{Wu16}, \cite{Liu17}, \cite{Zhang21}, \cite{Li22}. In \cite{Wu16} and \cite{Liu17}, the authors discover earlier the uncertainty in inference time and analyze the causes of inference time uncertainty. \cite{Wu16} evaluates the inference time performance of convolutional neural networks on multiple generations of iPhone SoC chips, observing significant performance variations through numerous outliers. The analysis shows that the inference time, particularly on the A11 chip, follows an approximately Gaussian distribution. \cite{Liu17} observes that the inference time of various DNN models applied to autonomous driving is uncertain, and the influence on the inference time fluctuation is analyzed from six aspects: data, I/O, model, runtime, hardware, and end-to-end perception system. Uncertainty in inference time brings a significant challenge to time-critical tasks. To address this challenge, \cite{Zhang21} designs a kernel-based prediction method to estimate DNN inference time on different devices, addressing the issue of not being able to obtain inference time \emph{a priori}. \cite{Li22} develops a method to estimate end-to-end inference time by training machine learning models to predict the time of each neural architecture component with limited profiling data and across different machine learning frameworks.

However, inference time exhibits significant randomness across different DNN models and hardware platforms, and the prediction methods proposed by \cite{Zhang21} and \cite{Li22} have not satisfied the requirements of high precision. The above studies do not involve the impact of computing resources on inference time and uncertainty, nor do they consider DNN partitioning decisions under inference time uncertainty. In this paper, our goal is to jointly optimize DNN model partitioning and the allocation of computing and communication resources to minimize energy consumption on mobile devices while satisfying probabilistic task deadlines. To the best of our knowledge, this issue has not been explored in the context of DNN partitioning.

\section{System Model and Problem Formulation}

\begin{figure}[t]
  \captionsetup{font={small}}
  \centering
  \includegraphics[width=0.5\textwidth]{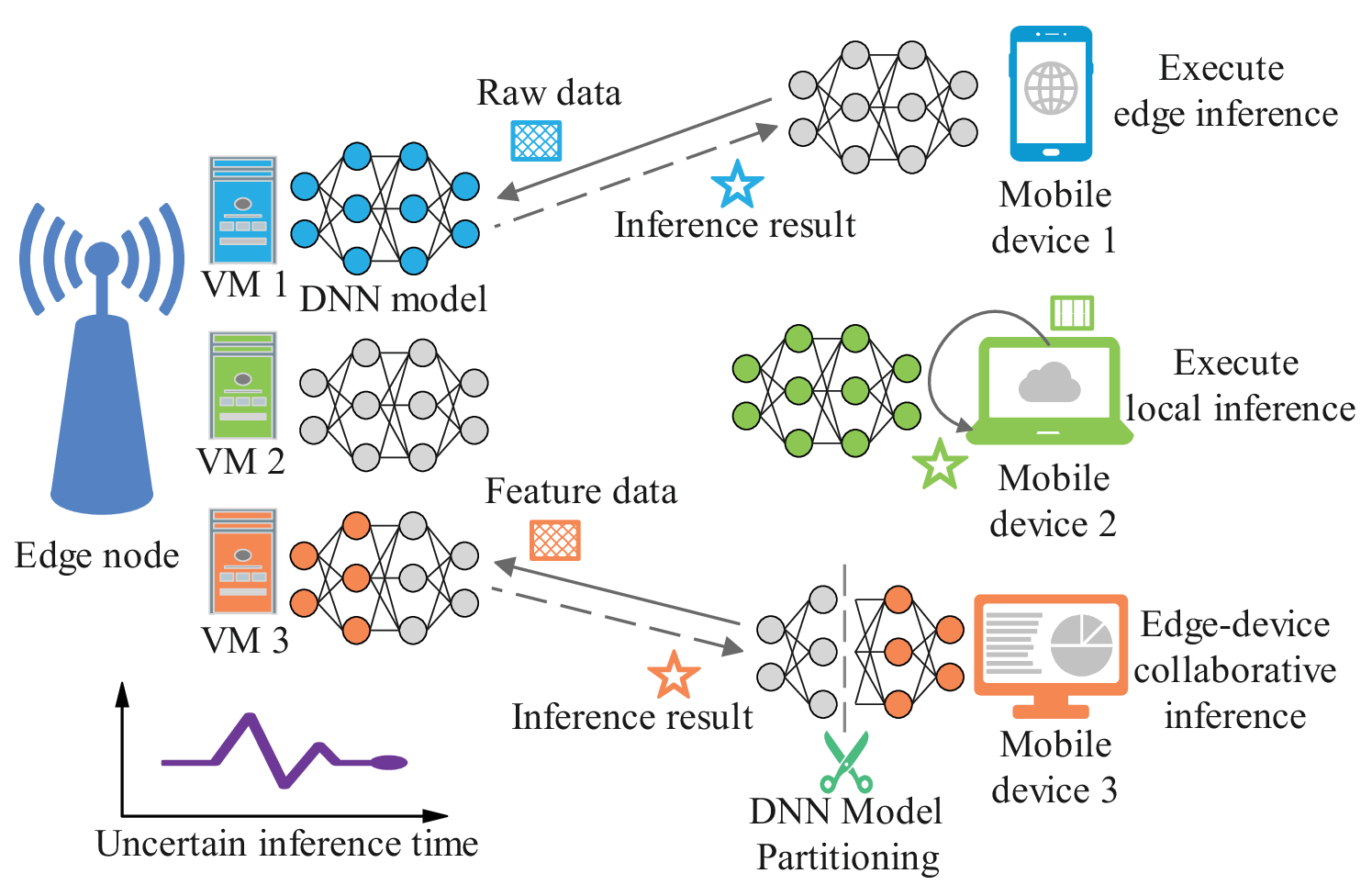}\\
  \caption{An example of the considered DNN model partitioning under inference time uncertainty in edge intelligence systems.}
  \label{fig:SysMo}
  \vspace{-0.1cm}
\end{figure}

\captionsetup[table]{labelsep=newline}
\begin{table*}[!htb]
    \footnotesize
    \caption{\label{tab:I}\textsc{Summary of Main Symbols}}
	\centering
	\begin{tabular}{p{1.0cm}<{\centering}|p{6.2cm}<{\centering}|p{1.0cm}<{\centering}|p{6.2cm}<{\centering}}
		\hline
        \textcolor{black}{\textbf{Symbol}} &\textcolor{black}{\textbf{Description} }
        & \textcolor{black}{\textbf{Symbol}} &\textcolor{black}{\textbf{Description} } \\
		\hline
		\textcolor{black}{$n$} &\textcolor{black}{Index of the $n$th mobile device}
		& \textcolor{black}{$\kappa_n$} & \textcolor{black}{Energy efficiency coefficient}  \\
		\hline
		\textcolor{black}{$\mathcal{N}$} &\textcolor{black}{Set of $N$ mobile devices}
		& \textcolor{black}{$B$} &\textcolor{black}{Total communication bandwidth}  \\
		\hline
		\textcolor{black}{$\mathcal{M}$} &\textcolor{black}{Set of $M$ partitioning points}
		& \textcolor{black}{$b_n$} &\textcolor{black}{Bandwidth allocated to mobile device $n$} \\
		\hline
		\textcolor{black}{$x_{n,m}$} &\textcolor{black}{Partitioning decision of mobile device $n$}
		& \textcolor{black}{$f_{\min}$} &\textcolor{black}{Minimum CPU/GPU frequency of the mobile device} \\
		\hline
		\textcolor{black}{$t_{n,m}^{\mathrm{loc}}$}& \textcolor{black}{Local inference time of mobile device $n$}
		& \textcolor{black}{$f_{\max}$} &\textcolor{black}{Maximum CPU/GPU frequency for the mobile device} \\
		\hline
		\textcolor{black}{$e_{n,m}^{\mathrm{loc}}$} &\textcolor{black}{Local energy consumption of mobile device $n$}
		& \textcolor{black}{$f_n$}  &\textcolor{black}{CPU/GPU frequency allocated to mobile device $n$} \\
		\hline
		\textcolor{black}{$t_{n, m}^{\mathrm{off}}$}& \textcolor{black}{Offloading time of mobile device $n$}
		& \textcolor{black}{$d_{n,m}$} &\textcolor{black}{Output data size by the $m$th block of the DNN model} \\
		\hline
		\textcolor{black}{$e_{n, m}^{\mathrm{off}}$} &\textcolor{black}{Offloading energy consumption of mobile device $n$}
		& \textcolor{black}{$p_{n}$} &\textcolor{black}{Transmission power of mobile device $n$} \\
		\hline		
		\textcolor{black}{$t_{n, m}^{\mathrm{vm}}$} &\textcolor{black}{Edge inference time of mobile device $n$}
		& \textcolor{black}{$h_{n}$} &\textcolor{black}{Channel gain of mobile device $n$} \\
		\hline
		\textcolor{black}{$D_n$} &\textcolor{black}{Deadline of the inference task}
	    & \textcolor{black}{$N_0$} &\textcolor{black}{Noise power spectral density} \\
		\hline
		\hline
	\end{tabular}
\end{table*}

As illustrated in Fig. \ref{fig:SysMo}, we consider a multi-device edge intelligence system consisting of $N$ mobile devices, represented by the set $\mathcal{N} \triangleq\{1, \ldots, N\}$, and one edge node integrated with an MEC server, where the mobile devices and the edge node only have one single antenna. The Frequency Division Multiple Access (FDMA) system is considered, where the channel interference between mobile devices can be negligible. We consider that each mobile device possesses a DNN model (e.g., AlexNet \cite{Krizhevsky23}, ResNet \cite{He24}, or VGG \cite{Simonyan25}) that can handle a certain number of inference tasks (e.g., image recognition). Meanwhile, the DNN model of each mobile device has an identical backup stored on the edge node.

\begin{figure}[t]
  \captionsetup{font={small}}
  \centering
  \includegraphics[width=0.5\textwidth]{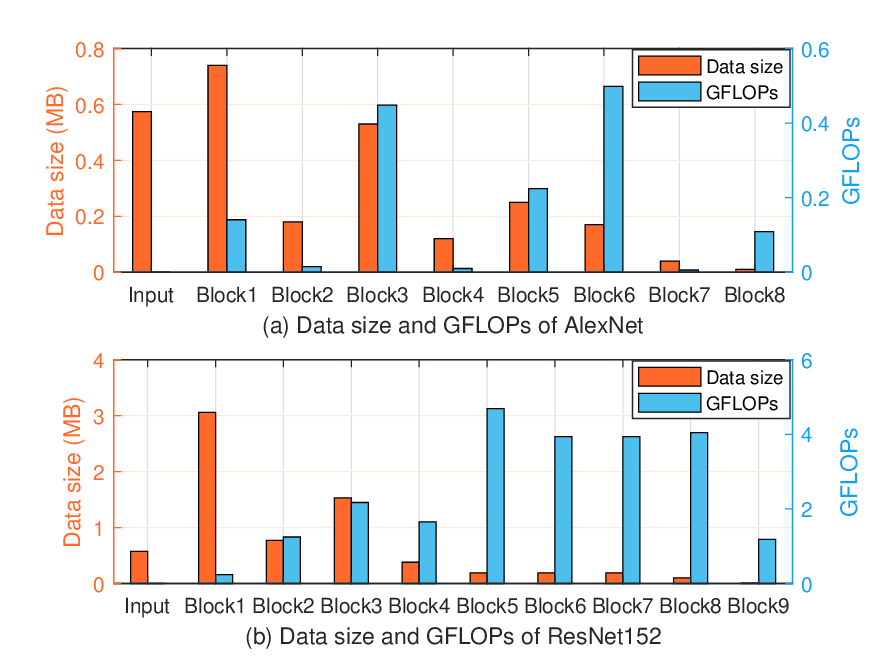}\\
  \caption{The data size and required GFLOPs of each block in AlexNet and ResNet152.}
  \label{fig:Data&FLOPs}
  \vspace{-0.1cm}
\end{figure}

In DNNs, the size of the output data (i.e., feature data) from some intermediate layers or blocks is typically smaller than the size of the input data (i.e., raw data). As the number of layers or blocks increases, the required computing capacity (i.e., GFLOPs) gradually rises. As shown in Fig. \ref{fig:Data&FLOPs}, the input data size of AlexNet and ResNet152 are both 0.574 MB. The feature data size of AlexNet's block $2$ and ResNet152's block $5$ are 0.18 MB and 0.19 MB, representing $69\%$ and $67\%$ reductions compared to the input data size. Correspondingly, after block 2, AlexNet requires GFLOPs that account for $90\%$ of the total GFLOPs, whereas ResNet152 needs $81\%$ of its total GFLOPs after block $5$. Therefore, inference tasks generated by resource-limited mobile devices can be offloaded to the MEC server with a powerful computing capacity for processing. More specifically, we can execute a part of the DNN inference task locally on the mobile device, offload a small amount of intermediate feature data to the MEC server, and then execute the remaining DNN inference task. The partitioning of DNN models needs to consider the tradeoff between computation and communication. From a more practical perspective, our work addresses the policy of DNN model partitioning and resource allocation when the inference time is not precisely known in advance. For ease of reference, the main symbols are summarized as Table I.

\subsection{DNN Model Partitioning}
\begin{figure}[t]
  \captionsetup{font={small}}
  \centering
  \includegraphics[width=0.50\textwidth]{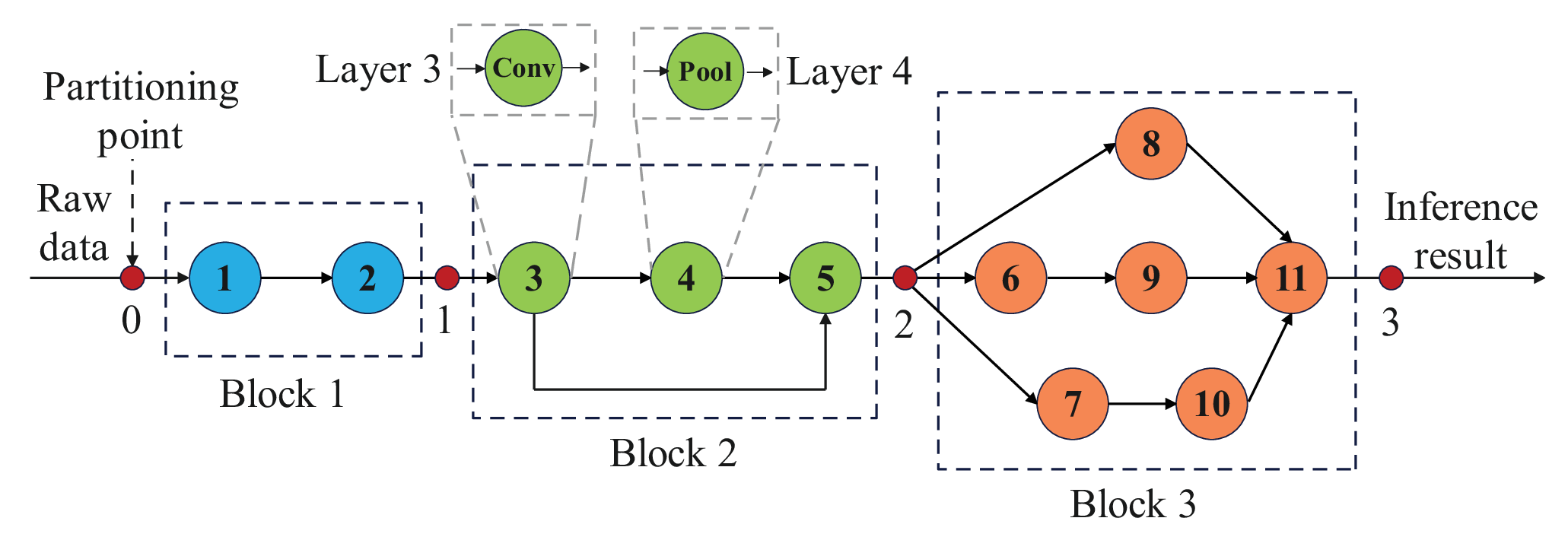}\\
  \caption{An example of the block-based DNN modeling and its partitioning points.}
  \label{fig:PartMo}
  \vspace{-0.1cm}
\end{figure}

Different DNNs exhibit a range of structures. For example, AlexNet and VGG are organized as single chains \cite{Krizhevsky23}, \cite{Simonyan25}, while ResNet features two asymmetric branches \cite{He24}. Typically, the structure of DNNs is modeled as DAGs \cite{Li14}, \cite{Liang26}. However, this DAG-based modeling can be quite complex. For simplicity, we use the block-based modeling approach \cite{Zhang10}, \cite{Shi12}. This method involves dividing the DAG into multiple blocks, effectively transforming it into a serial chain structure. As shown in Fig. \ref{fig:PartMo}, each block we construct consists of multiple layers, including convolutional layers (Conv), pooling layers (Pool), batch normalization layers (BN), activation layers (such as ReLU), etc. Denote $M$ as the number of blocks in the DNN model. Then, the set of partitioning points is represented as $\mathcal{M} \triangleq\{0, 1, \ldots, M\}$. Let $x_{n,m} \in\{0,1\}, n \in \mathcal{N}, m \in \mathcal{M}$ be the partitioning decision, and there is only one partitioning point for each mobile device, i.e., $\sum_{m \in \mathcal{M}} x_{n, m}=1, \forall n \in \mathcal{N}$. Specifically, $x_{n,m}=1$ indicates that mobile device $n$ executes partitioning at the $m$th point, and $x_{n,m}=0$ otherwise. For instance, $x_{n,0}=1$ means that mobile device $n$ only executes edge inference, $x_{n,M}=1$ means that mobile device $n$ only executes local inference, and $x_{n,m}=1$ means that the first $m$ blocks execute local inference, and the remaining $(M-m)$ blocks execute edge inference.

\subsection{Inference Time and Energy Consumption}
\begin{figure}[t]
  \captionsetup{font={small}}
  \centering
  \includegraphics[width=0.50\textwidth]{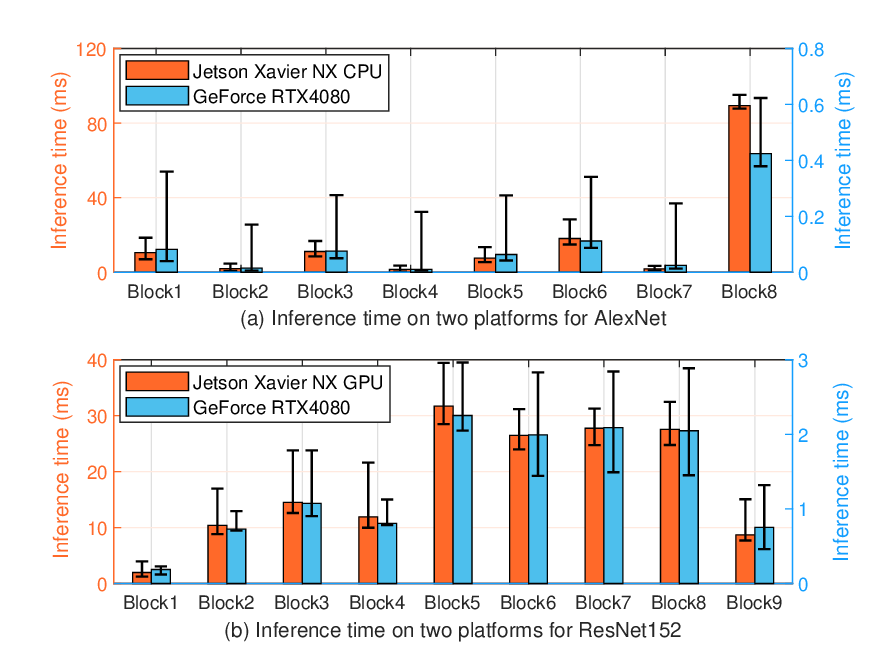}\\
  \caption{The variations in inference time on different platforms of each block for AlexNet and ResNet152.}
  \label{fig:BlockVar}
  \vspace{-0.1cm}
\end{figure}

As shown in Fig. \ref{fig:BlockVar}, the inference time of each block of AlexNet and ResNet152 on different hardware platforms is tested. The inference time of each block exhibits significant uncertainty and randomness, making it challenging to predict and understand the distribution of inference time precisely. However, it is pleasing that on the higher-computing platform (i.e., GeForce RTX 4080), the inference time and variation for each block of AlexNet and ResNet152 are significantly reduced compared to the lower-computing platform (i.e., Jetson Xavier NX CPU/GPU). Therefore, dynamic voltage and frequency scaling (DVFS) can be employed to optimize local computing resource allocation on mobile devices, while task offloading can be used to transfer computing to the MEC server, thereby reducing both inference time and its variation.

We assume that the partitioning point is $m \in \mathcal{M}$, and then the partitioning point set $\mathcal{M}$ is divided into two mutually exclusive sets $\mathcal{M}_0 \triangleq \{0,1, \ldots, m\}$ and $\mathcal{M}_1 \triangleq \mathcal{M} \backslash \mathcal{M}_0 \triangleq \{m+1, \ldots, M\}$. Let $u_{n, k}^{\mathrm{loc}}$ denote the local inference time of the mobile device $n$ in the $k$th block. Then, the local inference time of mobile device $n$ can be written as
\begin{equation}
t_{n, m}^{\mathrm{loc}}=\sum_{k \in \mathcal{M}_0} u_{n, k}^{\mathrm{loc}}, \forall n \in \mathcal{N}, m \in \mathcal{M},
\end{equation}
where $t_{n, 0}^{\mathrm{loc}}=u_{n, 0}^{\mathrm{loc}}=0$. The dynamic power consumption of the COMS circuit is denoted as $\alpha c V^2 f$, where $\alpha$ is the activity factor, $c$ is the load capacitance, $V$ is the supply voltage, and $f$ is the CPU/GPU clock frequency \cite{Yahya27}. Moreover, $V$ is approximately linear to the frequency when the CPU/GPU operates in the non-low frequency range, i.e., $V=kf$ \cite{Yahya27}, \cite{Wen28}. Thus, the corresponding energy consumption of mobile device $n$ to execute local inference is
\begin{equation}\label{eq:LocEC}
e_{n, m}^{\mathrm{loc}}=\kappa_n f_n^3 t_{n, m}^{\mathrm{loc}}, \forall n \in \mathcal{N}, m \in \mathcal{M},
\end{equation}
where $\kappa_n = \alpha_n c_n k_n^2$ is an energy efficiency coefficient that depends on the chip architecture.

Let $b_{n}$ denote the uplink bandwidth allocated by the edge node to mobile device $n$ for edge inference. The uplink bandwidth allocated to each mobile device is constrained by total bandwidth resource $B$, i.e., $\sum_{n \in \mathcal{N}} b_n \leq B$. The spectral efficiency of wireless uplink between the edge node and mobile device $n$ is $\eta_n^{\mathrm{off}}=\log _2\left(1+p_n h_n /b_n N_0\right)$, where $p_n$ is the transmission power, $h_n$ is the channel gain, and $N_0$ is the noise power spectral density. The offloading time of mobile device $n$ to transmit data to the edge node can be given as
\begin{equation}
t_{n, m}^{\text {off}}=\frac{d_{n, m}}{b_n \eta_n^{\mathrm{off}}}, \forall n \in \mathcal{N}, m \in \mathcal{M},
\end{equation}
where $d_{n, m}$ is the output data size by the $m$th block of the DNN model of mobile device $n$. Based on the partitioning decision $x_{n, m}$, $d_{n, m}$ can represent the size of the raw data, feature data, or result data. For instance, $d_{n, 0}$ denotes the size of the raw data, while $d_{n, M}$ represents the size of the result data. The corresponding offloading energy consumption of mobile device $n$ is
\begin{equation}
e_{n, m}^{\text {off }}=\frac{p_n d_{n, m}}{b_n \eta_n^{\mathrm{off}}}, \forall n \in \mathcal{N}, m \in \mathcal{M}.
\end{equation}

The MEC server can generate a virtual machine (VM) for each mobile device that executes edge inference. Each VM is configured with the corresponding DNN model to its associated mobile device and performs the offloading task in parallel. For simplicity, we assume that the total computing resources on the edge server are equally allocated to each VM, and each VM is dedicated to serving its corresponding mobile device \cite{Liang46}, \cite{Yan47}. The delay introduced by VM configuration is usually small (e.g., $\sim 10 \mathrm{~ms}$), so its impact on the overall delay is negligible and has been disregarded.\footnote{For example, advanced VM configuration techniques can make its latency less than 10 ms, and techniques such as hot start and lightweight configuration can further reduce the latency of VM configuration \cite{Long48}, \cite{Li49}.} Let $u_{n, k}^{\mathrm{vm}}$ denote the edge inference time of mobile device $n$ in the $k$th block. The edge inference time of mobile device $n$ can be expressed as
\begin{equation}
t_{n, m}^{\mathrm{vm}}=\sum_{k \in \mathcal{M}_1} u_{n, k}^{\mathrm{vm}}, \forall n \in \mathcal{N}, m \in \mathcal{M},
\end{equation}
where $t_{n, M}^{\mathrm{vm}}=0$. The time spent downloading the task computation results from the edge node (e.g., 5G Base Station) to the device is negligible, due to the strong transmission power and high bandwidth of the edge node, as well as the relatively small output data size (compared to the raw or feature data size). For example, in object recognition applications, the output typically consists of vehicles, pedestrians, etc., which requires only a few bytes and is much smaller than the corresponding input raw data or feature data, the latter of which may be several megabytes. In addition, since the MEC server is powered by the grid, the energy consumption of edge inference and result downloading is not considered \cite{Chen29}, \cite{Nan30}.

\subsection{Problem Formulation}
From the above analysis, the energy consumption of mobile device $n$ is
\begin{equation}
E_n=\sum_{m \in \mathcal{M}} x_{n, m}\left(e_{n, m}^{\mathrm{loc}}+e_{n, m}^{\mathrm{off}}\right), \forall n \in \mathcal{N}.
\end{equation}
Meanwhile, the inference time of mobile device $n$ is
\begin{equation}
T_n=\sum_{m \in \mathcal{M}} x_{n, m}\left(t_{n, m}^{\mathrm{loc}}+t_{n, m}^{\mathrm{off}}+t_{n, m}^{\mathrm{vm}}\right), \forall n \in \mathcal{N}.
\end{equation}

Due to the uncertainty of inference time, the actual inference time $T_n$ of mobile device $n$ is a random variable. Consequently, we would like to provide a probabilistic guarantee for the inference task with a hard deadline constraint under uncertainty of inference time, which is given as follows
\begin{equation}\label{eq:ChanceConstr}
\mathbb{P}\left\{T_n \leq D_n\right\} \geq 1-\varepsilon_n, \forall n \in \mathcal{N},
\end{equation}
where $D_n$ is the deadline of the inference task, and $\varepsilon_n$ is the violation probability that mobile device $n$ can tolerate, which is a small positive constant also called \emph{risk level}. In robust optimization, constraint (\ref{eq:ChanceConstr}) is generally called the chance constraint \cite{Nemirovski34}, \cite{Ben-Tal35}.

The objective is to jointly optimize DNN partitioning decision $\mathbf{x} \triangleq\left\{x_{n, m}\right\}_{n \in \mathcal{N}, m \in \mathcal{M}}$, uplink bandwidth allocation $\mathbf{b} \triangleq\left\{b_n\right\}_{n \in \mathcal{N}}$, and local computing resource allocation $\mathbf{f} \triangleq\left\{f_n\right\}_{n \in \mathcal{N}}$ to minimize the expected energy consumption of all mobile devices while satisfying the chance constraints. The optimization problem is formulated as
\begin{subequations}\label{eq:Problem1}
\begin{alignat}{2}
& \min _{\mathbf{x}, \mathbf{b}, \mathbf{f}} \mathbb{E}\left[\sum_{n \in \mathcal{N}} E_n\right] \\
& \ \text {s.t.} \ \, \mathbb{P}\left\{T_n \leq D_n\right\} \geq 1-\varepsilon_n, \forall n \in \mathcal{N}, \\
& \ \quad \  \sum_{m \in \mathcal{M}}x_{n, m}=1, \forall n \in \mathcal{N}, \\
& \ \quad \  \sum_{n \in \mathcal{N}} \sum_{m \in \mathcal{M}} x_{n, m} b_n \leq B, \\
& \ \quad \ \, x_{n, m} \in\{0,1\}, \forall n \in \mathcal{N}, m \in \mathcal{M}, \\
& \ \quad \ \, b_n \geq 0, \forall n \in \mathcal{N}, \\
& \ \quad \ \, f_{\min} \leq f_n \leq f_{\max}, \forall n \in \mathcal{N},
\end{alignat}
\end{subequations}
where (\ref{eq:Problem1}b) corresponds to the guarantee of chance constraints for hard deadlines under uncertain inference time, (\ref{eq:Problem1}c) and (\ref{eq:Problem1}e) correspond to constraints on DNN partitioning decisions, (\ref{eq:Problem1}d) represents constraints on uplink bandwidth allocation, and (\ref{eq:Problem1}f) and (\ref{eq:Problem1}g) indicate uplink bandwidth and local computing resources that can be allocated to mobile devices, respectively.

Although problem (\ref{eq:Problem1}) is easy to understand, solving it in practice is quite challenging. First and foremost, constraint (\ref{eq:Problem1}b) indicates the need to provide chance constraints for the inference of each task with a hard deadline, which is difficult to handle. Second, similar to \cite{Shi12}, \cite{Su13} and \cite{Zhang36}, the corresponding DNN partitioning and resource allocation remains a mixed-integer non-linear programming (MINLP) problem even given the deterministic inference time, which is generally NP-hard. To address above challenges, in the absence of precise inference time and its complete distributional knowledge, we develop a robust DNN model partitioning and resource allocation policy that relies solely on the mean and variance information of the inference time. The specific solutions are presented in Section IV and Section V.

\emph{Remark 1:} It is worth noting that the problem (\ref{eq:Problem1}) we propose can be simplified to the case of the previous work by setting the risk level of each mobile device to zero, and the mean and variance of the inference time for each block to true and zero, respectively. In this regard, the problem of uncertain inference time explored in this paper is both more meaningful and more challenging.

\section{Mean and Variance of Inference Time with Frequency Scaling}
In this section, we first provide a fitting function of relationship between mean inference time and CPU/GPU frequency using nonlinear least squares method. Then, we present an efficient method to estimate the variance and covariance of inference time across different CPU/GPU frequencies.

\subsection{Mean Inference Time}
The DVFS technology can be used to optimize inference time and energy consumption. Therefore, it is essential to give a model that accurately characterizes the relationship between CPU/GPU frequency and inference time. Most existing work models the inference time as a function of the workload and the CPU/GPU frequency, typically expressed as their ratio. The specific model is defined as $t=\frac{w}{gf}$, where $w$ (in GFLOPs) is the workload of the task, $f$ (in GHz) is the CPU/GPU frequency, and $g$ (in FLOPs/cycle) is the workload it can process per cycle \cite{Zeng31}, \cite{Han32}. However, we find that the parameter $g$ in the above model varies across different DNNs and within different blocks of the same DNN. As illustrated in Fig. \ref{fig:Data&FLOPs} and Fig. \ref{fig:BlockVar}, the total inference time of different DNNs is not necessarily proportional to the total GFLOP under a fixed CPU/GPU frequency. Similarly, the inference time of each block within the same DNN does not necessarily scale proportionally with its respective GFLOP. For example, the inference time of ResNet152 is 6-fold that of AlexNet, while the required GFLOPs are 16-fold higher than those of AlexNet. For AlexNet, the inference time of block $8$ is higher than that of other blocks, yet the required data size and GFLOPs are quite small.

\begin{figure}[t]
  \captionsetup{font={small}}
  \centering
  \includegraphics[width=0.50\textwidth]{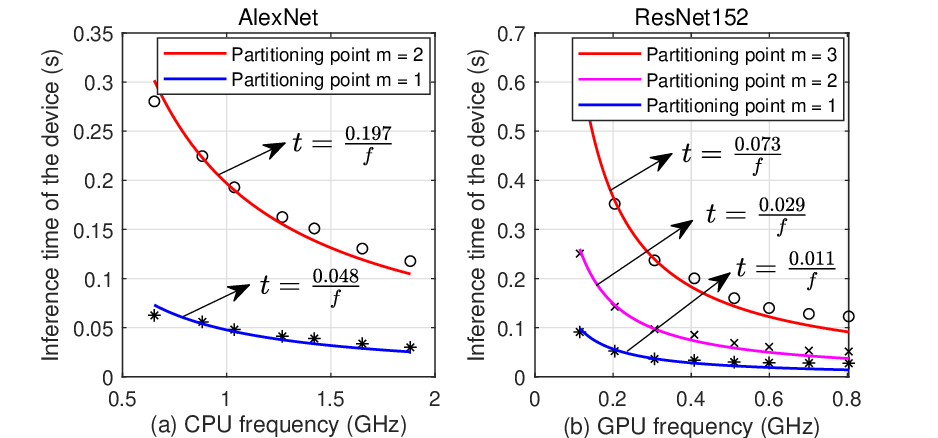}\\
  \caption{The mean inference time at different partitioning points on Jeston Xavier NX CPU/GPU. Discrete points represent the sample mean, while the continuous lines represent the fitted functions.}
  \label{fig:MeanFit}
  \vspace{-0.1cm}
\end{figure}

Therefore, we utilize real-world data to model the functional relationship between inference time and CPU/GPU frequency. The inference time of widely used DNNs (i.e., AlexNet and ResNet152) is tested on multiple devices (e.g., Jeston Xavier NX CPU and GPU) by frequency scaling. Specifically, AlexNet and ResNet152 are partitioned into $2$ and $3$ blocks, respectively. The sets of partitioning points are defined as $m \in M \triangleq\{0,1,2\}$ for AlexNet and $m \in M \triangleq\{0,1,2,3\}$ for ResNet152, where $m=0$ indicates that the inference is executed on the VM, resulting in the inference time of the device being $0$.\footnote{Due to space limitations, only the cases with 2 and 3 blocks are presented here. However, in the experiments where the blocks are partitioned into 8 or 9, each block demonstrated a similar curve, as depicted in Fig. 6.} We use nonlinear least square method to fit the measured data above. Fig. \ref{fig:MeanFit} illustrates the fitting curve and coefficient of AlexNet and ResNet152 for different partitioning points on CPU and GPU. For AlexNet, the squared 2-norm of the residual at $m=1$ and $m=2$ is 2.0e-4 $\text{s}^2$ and 9.7e-4 $\text{s}^2$, respectively. For ResNet152, the squared 2-norm of the residual at $m=1$, $m=2$, and $m=3$ is 5.7e-4 $\text{s}^2$, 8.0e-4 $\text{s}^2$, and 2.9e-3 $\text{s}^2$, respectively.

According to the above results, the mean inference time when mobile device $n$ selects partitioning point $m$ is modeled as follows:
\begin{equation}\label{eq:MeanFit}
\bar{t}_{n, m}^{\text {loc }}=\frac{w_{n, m}}{g_{n, m} f_n}, \forall n \in \mathcal{N}, m \in \mathcal{M},
\end{equation}
where $w_{n, m}$ is the GFLOPs required for local inference, $f_{n}$ is the local CPU/GPU frequency of mobile device $n$, and $g_{n, m}$ is the FLOPs that can be processed per cycle, which is decided by the partitioning point, the DNN model, and the CPU/GPU hardware.

\subsection{Variance and Covariance of Inference Time}
\begin{figure}[t]
  \captionsetup{font={small}}
  \centering
  \includegraphics[width=0.50\textwidth]{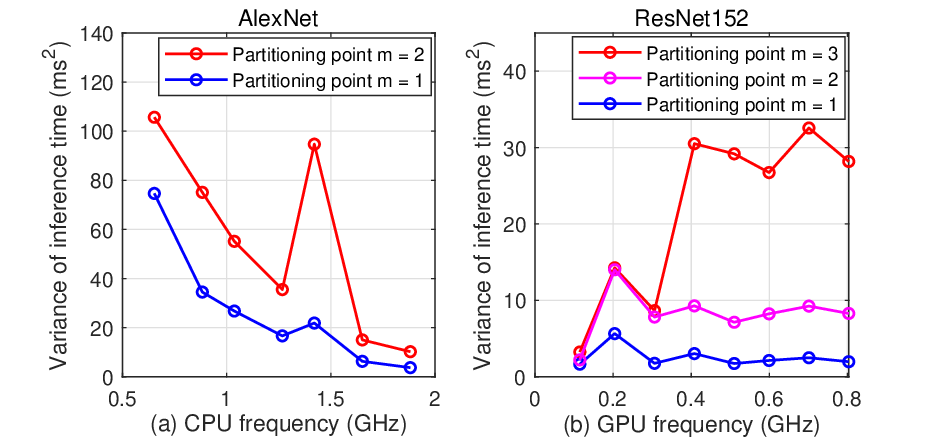}\\
  \caption{The variance of inference time at different partitioning points on Jeston Xavier NX CPU/GPU.}
  \label{fig:VarFit}
  \vspace{-0.1cm}
\end{figure}

Based on the measurement of mean inference time, the variance of inference time for AlexNet and ResNet152 at different CPU and GPU frequencies is calculated, as shown in Fig. \ref{fig:VarFit}. It can be observed that the variance of AlexNet is higher at low CPU frequencies, while the maximum variance of ResNet152 occurs at around $0.7$ GHz on the GPU. The results indicate that the variance of inference time is not a monotonic function of CPU/GPU frequency. In addition, compared to inference on the CPU, the variance of inference time on the GPU is relatively lower. However, the variance of inference time exhibits random and irregular fluctuations in response to variations in CPU/GPU frequency. Therefore, it is difficult to fit the relationship between variance and CPU/GPU frequency as a function, in contrast to the modeling of the mean inference time.

To solve the above problem, we use the mean value in the CPU/GPU frequency scaling range as the variance of the inference time. The variance of inference time when mobile device $n$ selects partitioning point $m$ is obtain by
\begin{equation}\label{eq:VarFit}
v_{n, m}^{\mathrm{loc}}=\frac{1}{|\mathcal{F}|}\sum_{\forall f_n \in \mathcal{F}} v_{n, m}^{\text {loc}}\left(f_n\right), \forall n \in \mathcal{N}, m \in \mathcal{M},
\end{equation}
where $v_{n, m}^{\text {loc}}(\cdot)=\mathbb{E}\left[\left(t_{n, m}^{\text {loc}}(\cdot)-\bar{t}_{n, m}^{\text {loc}}(\cdot)\right)^2\right]$ and $\mathcal{F}$ is the set of measurable frequency points of the CPU/GPU. This approximation may introduce some errors; however, simulation results show the error is acceptable. The experiments and analysis are discussed in Section VI. Note that in this work, we assume the CPU/GPU frequencies of mobile devices can be scaled, whereas the CPU/GPU frequencies of VMs remain constant. Therefore, $\overline{t}_{n, m}^{\mathrm{vm}}$ and $v_{n, m}^{\mathrm{vm}}$ can be obtained through simple online measurement.

During collaborative inference, covariance information between the mobile device and the VM is also required. Thus, we designate Jetson Xavier and Nano as the mobile device and RTX4080 as the VM, and calculate the covariance at different partitioning points. The experimental results show that the covariance curve closely matches the variance curve in Fig. \ref{fig:VarFit}. It is because the computing capacity of the VM is higher than mobile devices, leading to lower inference time and fluctuations. Therefore, similar to variance, the covariance of inference time at different partitioning points is approximated by
\begin{equation}\label{eq:CoVarFit}
w_{n, m, m^{\prime}}=\frac{1}{|\mathcal{F}|} \sum_{\forall f_n \in \mathcal{F}} w_{n, m, m^{\prime}}\left(f_n\right), \forall n \in \mathcal{N}, m, m^{\prime} \in \mathcal{M},
\end{equation}
where $w_{n, m, m^{\prime}}(\cdot)=\mathbb{E}\left[t_{n, m}(\cdot) t_{n, m^{\prime}}(\cdot)\right]-\bar{t}_{n, m}(\cdot) \bar{t}_{n, m^{\prime}}(\cdot)$.

\section{Robust DNN Partitioning and Resource Allocation}
To tackle the challenges posed by the chance constraints and combinatorial complexity of problem (\ref{eq:Problem1}), we first decompose problem (\ref{eq:Problem1}) into two subproblems: resource allocation and DNN model partitioning. Subsequently, the two subproblems with probabilistic constraints are equivalently transformed into deterministic optimization problems using the CCP method. Finally, convex optimization technique and PCCP technique are applied to obtain the optimal solution of the resource allocation subproblem and a stationary point of the DNN model partitioning subproblem respectively.

\captionsetup{font={small}}
\begin{figure*}[t]
\centering
\includegraphics[width=0.85\textwidth]{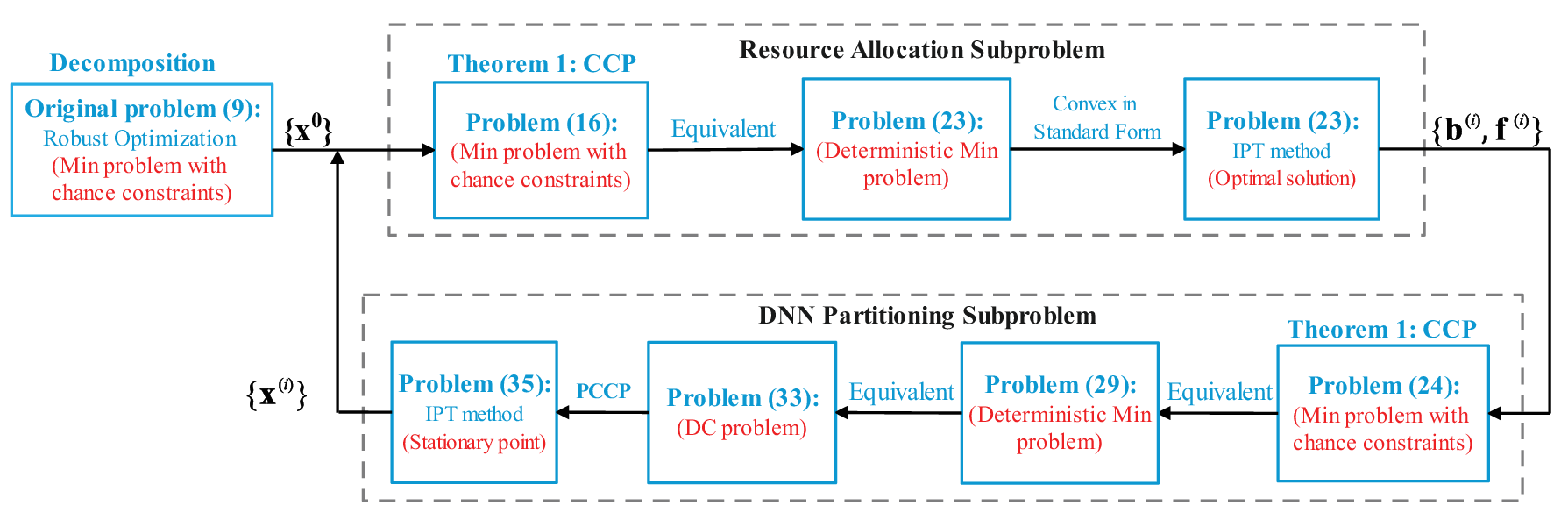}\\
\caption{\leftline{The general schematic of the optimization problem and corresponding solution.}}
\label{fig:FlowChart}
\vspace{-0.3cm}
\end{figure*}

\subsection{Problem Decomposition}
By leveraging the structure of the objective function and constraints in problem (\ref{eq:Problem1}), we find that it can be decomposed into two subproblems with separated objectives and constraints. We use the Tammer decomposition method \cite{Tammer37} to transform the high-complexity original problem into two lower-complexity subproblems and solve these subproblems alternately.
First, the resource allocation subproblem is written as
\begin{equation}\label{eq:Problem11}
\begin{aligned}
& \min _{\mathbf{b}, \mathbf{f}} E(\mathbf{b}, \mathbf{f} \mid \mathbf{x}) \\
& \ \ \text {s.t.} (\ref{eq:Problem1}\mathrm{b}), (\ref{eq:Problem1}\mathrm{d}), (\ref{eq:Problem1}\mathrm{f}), (\ref{eq:Problem1}\mathrm{g}).
\end{aligned}
\end{equation}
where $E(\mathbf{b}, \mathbf{f} \mid \mathbf{x})$ is the optimal value function corresponding to the resource allocation subproblem. Then, the DNN model partitioning subproblem is expressed as
\begin{equation}\label{eq:Problem12}
\begin{aligned}
& \min _{\mathbf{x}} E(\mathbf{x} \mid \mathbf{b}, \mathbf{f}) \\
& \ \ \text {s.t.} (\ref{eq:Problem1}\mathrm{b}), (\ref{eq:Problem1}\mathrm{c}), (\ref{eq:Problem1}\mathrm{d}), (\ref{eq:Problem1}\mathrm{e}).
\end{aligned}
\end{equation}
where $E(\mathbf{x} \mid \mathbf{b}, \mathbf{f})$ is the optimal value function corresponding to the DNN model partitioning subproblem. Note that the decomposition from the original problem (\ref{eq:Problem1}) to problem (\ref{eq:Problem11}) and problem (\ref{eq:Problem12}) does not change the optimality of the solution \cite{Tammer37}. In the following, we will give solutions of the resource allocation subproblem and the DNN model partitioning subproblem. The general schematic of the solution is shown in Fig. \ref{fig:FlowChart}.

\subsection{Resource Allocation Subproblem}
We define the set $\mathcal{G}$ that contains the partitioning points for all mobile devices as $\mathcal{G} \triangleq\left\{m_n \in \mathcal{M} \mid x_{n, m_n}=1, \forall n \in \mathcal{N}\right\}$. For a given DNN model partitioning decision $\mathbf{x} \triangleq\left\{x_{n, m_n}\right\}_{n \in \mathcal{N}, m_n \in \mathcal{G}}$, the expected energy consumption of mobile devices is
\begin{equation}
\mathbb{E}\left[\sum_{n \in \mathcal{N}} E_n\right]=\sum_{n \in \mathcal{N}}\left(\kappa_n \frac{w_{n, m_n}}{g_{n, m_n}} f_n^2+\frac{p_n d_{n, m_n}}{b_n \eta_n^{\mathrm{off}}}\right).
\end{equation}
Then, problem (\ref{eq:Problem11}) is rewritten as
\begin{subequations}\label{eq:RA-SubProblem0}
\begin{alignat}{2}
& \min _{\mathbf{b}, \mathbf{f}} \sum_{n \in \mathcal{N}}\left(\kappa_n \frac{w_{n, m_n}}{g_{n, m_n}} f_n^2 + \frac{p_n d_{n, m_n}}{b_n \eta_n^{\mathrm{off}}}\right) \\
& \ \ \text {s.t.} \  \mathbb{P}\left\{t_{n, m_n} \leq D_n\right\} \geq 1-\varepsilon_n, \forall n \in \mathcal{N}, m_n \in \mathcal{G}, \\
& \ \ \quad \ \sum_{n \in \mathcal{N}} b_n \leq B, \\
& \ \ \quad \ \, b_n \geq 0, \forall n \in \mathcal{N}, \\
& \ \ \quad \ \, f_{\min} \leq f_n \leq f_{\max}, \forall n \in \mathcal{N},
\end{alignat}
\end{subequations}
where $t_{n, m_n} \triangleq t_{n, m_n}^{\mathrm{loc}}+t_{n, m_n}^{\mathrm{off}}+t_{n, m_n}^{\mathrm{vm}}$ is the total inference time of mobile device $n$.

Due to the lack of the distribution of inference time, a difficult step is to reformulate the intractable chance constraints in (\ref{eq:RA-SubProblem0}b) into the deterministic constraints. To address this, we introduce a novel CCP technique \cite{Li38}, which does not introduce any relaxation in the optimization space when the chance constraint is transformed into a deterministic constraint. It allows that the mean and covariance of random variables can be measured without any assumptions. The details are given as follows:

\emph{Theorem 1:} Given random variables $\boldsymbol{\mathrm{\lambda}} \triangleq \left[\lambda_1, \lambda_2, \ldots, \lambda_n\right]^{\mathrm{T}}$ with known mean $\boldsymbol{\overline{\mathrm{\lambda}}}\triangleq \left[\overline{\lambda}_1, \overline{\lambda}_2, \ldots, \overline{\lambda}_n\right]^{\mathrm{T}}$ and covariance matrix $\boldsymbol{\mathrm{C}} \triangleq \mathbb{E}\left[(\boldsymbol{\mathrm{\lambda}}-\boldsymbol{\overline{\mathrm{\lambda}}})(\boldsymbol{\mathrm{\lambda}}-\boldsymbol{\overline{\mathrm{\lambda}}})^{\mathrm{T}}\right]$, a deterministic vector $\boldsymbol{\mathrm{a}}\triangleq \left[a_1, a_2, \cdots, a_n\right]^{\mathrm{T}}$, a constant $z$, and the risk level $\epsilon$, we can have the standard form of the \emph{Exact Conic Reformulation} (ECR) as follows
\begin{equation}
\mathbb{P}_{\boldsymbol{\lambda} \sim(\boldsymbol{\overline{\lambda}}, \boldsymbol{\mathrm{C}})}\left\{\boldsymbol{\mathrm{a}}^{\mathrm{T}} \boldsymbol{\lambda} \leq z\right\} \geq 1-\epsilon,
\end{equation}
if and only if
\begin{equation}
\boldsymbol{\mathrm{a}}^{\mathrm{T}} \overline{\boldsymbol{\mathrm{\lambda}}}+\sqrt{\frac{1-\epsilon}{\epsilon}} \sqrt{\boldsymbol{\mathrm{a}}^{\mathrm{T}} \boldsymbol{\mathrm{C}} \boldsymbol{\mathrm{a}}} \leq z,
\end{equation}
where $\boldsymbol{\mathrm{\lambda}} \sim(\boldsymbol{\overline{\mathrm{\lambda}}}, \boldsymbol{\mathrm{C}})$ indicates that the mean and covariance of the random variable $\boldsymbol{\mathrm{\lambda}}$ are $\boldsymbol{\overline{\lambda}}$ and $\boldsymbol{\mathrm{C}}$, respectively.

\begin{proof}
The proof of Theorem 1 is given in \cite{Li38}.
\end{proof}

Inspired by the CCP technique, we formulate the constraint (\ref{eq:RA-SubProblem0}b) into the standard form of the ECR, as follows:
\begin{equation}
\mathbb{P}_{\boldsymbol{\mathrm{\mu}}_n \sim(\boldsymbol{\overline{\mathrm{\mu}}}_n, \boldsymbol{\mathrm{V}}_n)}\left\{\boldsymbol{\mathrm{c}}_n^{\mathrm{T}} \boldsymbol{\mathrm{\mu}}_n \leq D_n\right\} \geq 1-\varepsilon_n, \forall n \in \mathcal{N},
\end{equation}
where $\boldsymbol{\mathrm{c}}_n^{\mathrm{T}} \triangleq \left[1, 1, 1\right]$ and $\boldsymbol{\mathrm{\mu}}_n \triangleq \left[t_{n, m_n}^{\mathrm{loc}}, t_{n, m_n}^{\mathrm{off}}, t_{n, m_n}^{\mathrm{vm}}\right]^{\mathrm{T}}$ for all $m_n \in \mathcal{G}$. The mean vector of $\boldsymbol{\mathrm{\mu}}_n$ is
\begin{equation}
\boldsymbol{\overline{\mathrm{\mathrm{\mu}}}}_n \triangleq \left[\overline{t}_{n, m_n}^{\mathrm{loc}}, \overline{t}_{n, m_n}^{\mathrm{off}}, \overline{t}_{n, m_n}^{\mathrm{vm}}\right]^{\mathrm{T}}, \forall n \in \mathcal{N}, m_n \in \mathcal{G},
\end{equation}
where $\overline{t}_{n, m_n}^{\mathrm{loc}}$ can be obtained by (\ref{eq:MeanFit}), $\overline{t}_{n, m_n}^{\mathrm{off}}=t_{n, m_n}^{\mathrm{off}}=d_{n, m_n} / b_n \eta_{n, m_n}^{\mathrm{off}}$ is the real offloading time.\footnote{This work does not consider channel state uncertainty and assumes that channel state information can be accurately obtained. However, our method can be extended to scenarios that jointly consider inference time and channel state uncertainty.}, $\overline{t}_{n, m_n}^{\mathrm{vm}}$ is the measured mean of $t_{n, m_n}^{\mathrm{vm}}$. Accordingly, the covariance matrix is constructed as
\begin{equation}
\boldsymbol{\mathrm{V}}_n \triangleq \left[\begin{array}{ccc}
v_{n, m_n}^{\mathrm{loc}} & 0 & 0 \\
0 & 0 & 0 \\
0 & 0 & v_{n, m_n}^{\mathrm{vm}}
\end{array}\right], \forall n \in \mathcal{N}, m_n \in \mathcal{G},
\end{equation}
where $v_{n, m_n}^{\mathrm{loc}}$ is given by (\ref{eq:VarFit}) and $v_{n, m_n}^{\mathrm{vm}}$ is the measured variances of $v_{n, m_n}^{\mathrm{vm}}$.

Based on Theorem 1, the chance constraints in (\ref{eq:RA-SubProblem0}b) with respect to bandwidth allocation $\mathbf{b}$ and computing resource allocation $\mathbf{f}$ are equivalently transformed to the following deterministic constraints:
\begin{align}\label{eq:RA-Determi}
\nonumber
& \left(\frac{w_{n, m_n}}{g_{n, m_n} f_n}+\frac{d_{n, m_n}}{b_n \eta_{n, m_n}^{\mathrm{off}}}+\overline{t}_{n, m_n}^{\mathrm{vm}}\right) + \\
& \qquad \sigma_n \sqrt{\left(v_{n, m_n}^{\mathrm{loc}}+v_{n, m_n}^{\mathrm{vm}}\right)} \leq D_n, \forall n \in \mathcal{N}, m_n \in \mathcal{G},
\end{align}
where $\sigma_n=\sqrt{\left(1-\varepsilon_n\right) / \varepsilon_n}$. After removing all random variables and considering $\boldsymbol{\overline{\mathrm{\mu}}_n}$ and $\boldsymbol{\mathrm{V_n}}$ as known constants, we derive an equivalent deterministic problem of problem (\ref{eq:RA-SubProblem0}) with the given DNN partitioning decision as follows:
\begin{subequations}\label{eq:RA-SubProblem1}
\begin{alignat}{2}
& \min _{\mathbf{b},\mathbf{f}} \sum_{n \in \mathcal{N}}\left(\kappa_n \frac{w_{n, m_n}}{g_{n, m_n}} f_n^2+\frac{p_n d_{n, m_n}}{b_n \eta_n^{\mathrm{off}}}\right) \\
& \ \ \text {s.t.} \ \, (\ref{eq:RA-SubProblem0}\mathrm{c}), (\ref{eq:RA-SubProblem0}\mathrm{d}), (\ref{eq:RA-SubProblem0}\mathrm{e}), (\ref{eq:RA-Determi}).
\end{alignat}
\end{subequations}
Note that problem (\ref{eq:RA-SubProblem1}) is convex, so the optimal resource allocation can be solved via an interior point (IPT) algorithm. The computational complexity of solving problem (\ref{eq:RA-SubProblem1}) using an IPT algorithm is $\mathcal{O}\left(N^3)\right)$, and the number of iterations of the IPT algorithm is $\mathcal{O}(\sqrt{N} \log (1/\xi))$, where $\xi$ is the convergence accuracy. Therefore, the total computational complexity is $\mathcal{O}\left(N^{3.5} \log (1/\xi)\right)$ \cite{Nesterov39}.

\subsection{DNN Model Partitioning Subproblem}
In the previous subsection, we obtained the optimal solution to the bandwidth allocation $\mathbf{b}$ and computing resource allocation $\mathbf{f}$ under a given $\mathbf{x}$. Next, we use the solutions $\mathbf{b}$ and $\mathbf{f}$ obtained from resource allocation subproblem (\ref{eq:RA-SubProblem0}) to optimize $\mathbf{x}$. The DNN model partitioning subproblem of problem (\ref{eq:Problem12}) can be rewritten as
\begin{subequations}\label{eq:Par-SubProblem0}
\begin{alignat}{2}
& \min _{\mathbf{x}} \sum_{n \in \mathcal{N}} \sum_{m \in \mathcal{M}} x_{n, m}\left(\kappa_n \frac{w_{n, m}}{g_{n, m}} f_n^2+\frac{p_n d_{n, m}}{b_n \eta_n^{\mathrm{off}}}\right) \\
& \ \ \text {s.t.} \ \, \mathbb{P}\left\{\sum_{m \in \mathcal{M}} x_{n, m} t_{n, m} \leq D_n\right\} \geq 1-\varepsilon_n, \forall n \in \mathcal{N}, \\
& \ \ \quad \ \sum_{m \in \mathcal{M}} x_{n, m}=1, \forall n \in \mathcal{N}, \\
& \ \ \quad \ \sum_{n \in \mathcal{N}} \sum_{m \in \mathcal{M}} x_{n, m} b_n \leq B, \\
& \ \ \quad \ \, x_{n, m} \in\{0,1\}, \forall n \in \mathcal{N}, m \in \mathcal{M} \text {, }
\end{alignat}
\end{subequations}
where $t_{n, m} \triangleq t_{n, m}^{\mathrm{loc}}+t_{n, m}^{\mathrm{off}}+t_{n, m}^{\mathrm{vm}}$. Note that in addition to the intractable chance constraints in (\ref{eq:Par-SubProblem0}b), problem (\ref{eq:Par-SubProblem0}) is non-convex due to the binary variable $\mathbf{x}$. We first transform the chance constraints into equivalent deterministic constraints. The constraint (\ref{eq:Par-SubProblem0}b) can be formulated as
\begin{equation}
\mathbb{P}_{\boldsymbol{\mathrm{\tau}}_n \sim(\boldsymbol{\overline{\mathrm{\tau}}}_n, \boldsymbol{\mathrm{W}}_n)}\left\{\boldsymbol{\mathrm{x}}_n^{\mathrm{T}} \boldsymbol{\mathrm{\tau}}_n \leq D_n\right\} \geq 1-\varepsilon_n, \forall n \in \mathcal{N},
\end{equation}
where $\boldsymbol{\mathrm{x}}_n^{\mathrm{T}}\triangleq \left[x_{n, 0}, x_{n, 1}, \ldots, x_{n, M}\right]$ is the partitioning decision vector, and $\boldsymbol{\mathrm{\tau}}_n \triangleq \left[t_{n, 0}, t_{n, 1}, \ldots, t_{n, M}\right]^{\mathrm{T}}$ is the inference time vector. The mean vector of $\boldsymbol{\mathrm{\tau}}_n$ is
\begin{equation}
\boldsymbol{\overline{\mathrm{\tau}}}_n \triangleq \left[\overline{t}_{n, 0}, \overline{t}_{n, 1}, \ldots, \overline{t}_{n, M}\right]^{\mathrm{T}}, \forall n \in \mathcal{N},
\end{equation}
where $\overline{t}_{n, m} \triangleq \overline{t}_{n, m}^{\mathrm{loc}}+\overline{t}_{n, m}^{\mathrm{off}}+\overline{t}_{n, m}^{\mathrm{vm}}$ for all $m \in \mathcal{M}$. $\overline{t}_{n, m}^{\mathrm{loc}}$  can be given by (\ref{eq:MeanFit}), $\overline{t}_{n, m}^{\mathrm{off}}=t_{n, m}^{\mathrm{off}}=d_{n, m} / b_n \eta_{n, m}^{\mathrm{off}}$, and $\overline{t}_{n, m}^{\mathrm{vm}}$ is the measured mean of $t_{n, m}^{\mathrm{vm}}$. Consequently, the covariance matrix is defined as
\begin{equation}
\boldsymbol{\mathrm{W}_n}=\left[\begin{array}{cccc}
w_{n, 0,0} & w_{n, 0,1} & \cdots & w_{n, 0, M} \\
w_{n, 1,0} & w_{n, 1,1} & \cdots & w_{n, 1, M} \\
\vdots & \vdots & \vdots & \vdots \\
w_{n, M, 0} & w_{n, M, 1} & \cdots & w_{n, M, M}
\end{array}\right], \forall n \in \mathcal{N},
\end{equation}
where $w_{n, m, m^{\prime}}$ is obtained by (\ref{eq:CoVarFit}).

Based on Theorem 1, the chance constraints in (\ref{eq:Par-SubProblem0}b) with respect to DNN model partitioning $\mathbf{x}$ are equivalently transformed to the following deterministic constraints:
\begin{equation}\label{eq:Par-Determi}
\sum_{m \in \mathcal{M}} x_{n, m} \overline{t}_{n, m}+\sigma_n \sqrt{\sum_{m \in \mathcal{M}} w_{n, m, m} x_{n, m}^2} \leq D_n, \forall n \in \mathcal{N}.
\end{equation}
where $w_{n, m, m}$ is the diagonal element of the $\boldsymbol{\mathrm{W}_n}$ matrix. Then, we replace constraint (\ref{eq:Par-SubProblem0}b) in problem (\ref{eq:Par-SubProblem0}) with the constraint (\ref{eq:Par-Determi}), and reformulate problem (\ref{eq:Par-SubProblem0}) as an equivalent deterministic problem as follows:
\begin{subequations}\label{eq:Par-SubProblem1}
\begin{alignat}{2}
& \min _{\mathbf{x}} \sum_{n \in \mathcal{N}} \sum_{m \in \mathcal{M}} x_{n, m}\left(\kappa_n \frac{w_{n, m}}{g_{n, m}} f_n^2+\frac{p_n d_{n, m}}{b_n \eta_n^{\mathrm{off}}}\right) \\
& \ \ \text {s.t.} \ \, (\ref{eq:Par-SubProblem0}\mathrm{c}), (\ref{eq:Par-SubProblem0}\mathrm{d}), (\ref{eq:Par-SubProblem0}\mathrm{e}), (\ref{eq:Par-Determi}).
\end{alignat}
\end{subequations}

To handle the combinatorial nature of the binary variable $\mathbf{x}$ of problem (\ref{eq:Par-SubProblem1}), we transform problem (\ref{eq:Par-SubProblem1}) into an equivalent difference-of-convex (DC) problem and obtain a stationary point of problem (\ref{eq:Par-SubProblem1}) using the PCCP technique. In what follows, we first replace the binary constraints in (\ref{eq:Par-SubProblem0}e) with the following constraints:
\begin{equation}\label{eq:RelaxBin}
x_{n, m} \in[0,1], \forall n \in \mathcal{N}, m \in \mathcal{M},
\end{equation}
\begin{equation}
x_{n, m}\left(1-x_{n, m}\right) \leq 0, \forall n \in \mathcal{N}, m \in \mathcal{M},
\end{equation}
Then, we introduce auxiliary variables $\boldsymbol{\mathrm{y}} \triangleq\left\{y_n\right\}_{n \in \mathcal{N}}$:
\begin{equation}
y_n=\sqrt{\sum_{m \in \mathcal{M}} w_{n, m, m} x_{n, m}^2}, \forall n \in \mathcal{N},
\end{equation}
where $y_n>0$ for all $n \in \mathcal{N}$. Therefore, problem (\ref{eq:Par-SubProblem1}) can be equivalently transformed into the problem as follows:
\begin{subequations}\label{eq:Par-SubProblem2}
\begin{alignat}{2}
& \min _{\mathbf{x}, \mathbf{y}} \sum_{n \in \mathcal{N}} \sum_{m \in \mathcal{M}} x_{n, m}\left(\kappa_n \frac{w_{n, m}}{g_{n, m}} f_n^2+\frac{p_n d_{n, m}}{b_n \eta_n^{\mathrm{off}}}\right) \\
& \ \ \text {s.t.} \ \, (\ref{eq:Par-SubProblem0}\mathrm{c}), (\ref{eq:Par-SubProblem0}\mathrm{d}), (\ref{eq:RelaxBin}), \\
& \ \ \quad \ \sum_{m \in \mathcal{M}} x_{n, m} \bar{t}_{n, m}+\sigma_n y_n \leq D_n, \forall n \in \mathcal{N}, \\
& \ \ \quad \ \sum_{m \in \mathcal{M}} w_{n, m, m} x_{n, m}^2-y_n^2 \leq 0, \forall n \in \mathcal{N}, \\
& \ \ \quad \ \, y_n^2-\sum_{m \in \mathcal{M}} w_{n, m, m} x_{n, m}^2 \leq 0, \forall n \in \mathcal{N}, \\
& \ \ \quad \ \, x_{n, m}-x_{n, m}^2 \leq 0, \forall n \in \mathcal{N}, m \in \mathcal{M}, \\
& \ \ \quad \ \, y_n>0, \forall n \in \mathcal{N},
\end{alignat}
\end{subequations}
where objective function (\ref{eq:Par-SubProblem2}a), constraints (\ref{eq:Par-SubProblem2}b), (\ref{eq:Par-SubProblem2}c) and (\ref{eq:Par-SubProblem2}g) are convex. However, there are concave functions in constraints (\ref{eq:Par-SubProblem2}d), (\ref{eq:Par-SubProblem2}e) and (\ref{eq:Par-SubProblem2}f), for which problem (\ref{eq:Par-SubProblem2}) is identified as a DC problem that can be solved using the PCCP technique \cite{Lipp40}.

We first relax problem (\ref{eq:Par-SubProblem2}) by adding relaxation variables to the DC constraints and penalizing the sum of violations to avoid the infeasibility of each iteration.
The penalty function can be given as
\begin{equation}
P=\sum_{n \in \mathcal{N}} \alpha_n+\sum_{n \in \mathcal{N}} \beta_n+\sum_{n \in \mathcal{N}} \sum_{m \in \mathcal{M}} \gamma_{n, m},
\end{equation}
where $\boldsymbol{\alpha} \triangleq\left\{\alpha_n\right\}_{n \in \mathcal{N}}$, $\boldsymbol{\beta} \triangleq\left\{\beta_n\right\}_{n \in \mathcal{N}}$, and $\boldsymbol{\gamma} \triangleq\left\{\gamma_{n, m}\right\}_{n \in \mathcal{N}, m \in \mathcal{M}}$ are slack variables added for constraints (\ref{eq:Par-SubProblem2}d), (\ref{eq:Par-SubProblem2}e), and (\ref{eq:Par-SubProblem2}f), respectively. Accordingly, the penalty DC problem can be obtained as
\begin{subequations}\label{eq:Par-SubProblem3}
\begin{alignat}{2}
& \min _{\substack{\mathbf{x}, \mathbf{y} \\ \boldsymbol{\alpha}, \boldsymbol{\beta}, \boldsymbol{\gamma}}} \ \sum_{n \in \mathcal{N}} \sum_{m \in \mathcal{M}} x_{n, m}\left(\kappa_n \frac{w_{n, m}}{g_{n, m}} f_n^2 + \frac{p_n d_{n, m}}{b_n \eta_n^{\mathrm{off}}}\right)+ \rho P \\
& \ \ \text {s.t.} \ \, (\ref{eq:Par-SubProblem0}\mathrm{c}), (\ref{eq:Par-SubProblem0}\mathrm{d}), (\ref{eq:RelaxBin}), (\ref{eq:Par-SubProblem2}\mathrm{c}), (\ref{eq:Par-SubProblem2}\mathrm{g})\\
& \ \ \quad \ \sum_{m \in \mathcal{M}} w_{n, m, m} x_{n, m}^2-y_n^2 \leq \alpha_n, \forall n \in \mathcal{N}, \\
& \ \ \quad \ \, y_n^2-\sum_{m \in \mathcal{M}} w_{n, m, m} x_{n, m}^2 \leq \beta_n, \forall n \in \mathcal{N}, \\
& \ \ \quad \ \, x_{n, m}-x_{n, m}^2 \leq \gamma_{n, m}, \forall n \in \mathcal{N}, m \in \mathcal{M}, \\
& \ \ \quad \ \, \alpha_n \geq 0, \beta_n \geq 0, \gamma_{n, m} \geq 0, \forall n \in \mathcal{N}, m \in \mathcal{M},
\end{alignat}
\end{subequations}
where $\rho>0$ is a penalty parameter. Then, the concave terms of the constraints (\ref{eq:Par-SubProblem3}c), (\ref{eq:Par-SubProblem3}d), and (\ref{eq:Par-SubProblem3}e) are linearized to obtain convex constraints for a minimization problem and to solve a sequence of convex problems successively. Specifically, at $i$th iteration, update $\left\{\mathbf{x}^{(i)}, \mathbf{y}^{(i)}\right\}$ by solving the following approximate problem, which is parameterized by $\left\{\mathbf{x}^{(i-1)}, \mathbf{y}^{(i-1)}\right\}$ obtained at $(i-1)$th iteration.
\begin{subequations}\label{eq:Par-SubProblem4}
\begin{alignat}{2}
& \min _{\substack{\mathbf{x}, \mathbf{y} \\ \boldsymbol{\alpha}, \boldsymbol{\beta}, \boldsymbol{\gamma}}} \ \sum_{n \in \mathcal{N}} \sum_{m \in \mathcal{M}} x_{n, m}\left(\kappa_n \frac{w_{n, m}}{g_{n, m}} f_n^2 + \frac{p_n d_{n, m}}{b_n \eta_n^{\mathrm{off}}}\right)+ \rho^{(i-1)} P \\
& \text {s.t.} \ \, (\ref{eq:Par-SubProblem0}\mathrm{c}), (\ref{eq:Par-SubProblem0}\mathrm{d}), (\ref{eq:RelaxBin}), (\ref{eq:Par-SubProblem2}\mathrm{c}), (\ref{eq:Par-SubProblem2}\mathrm{g}), (\ref{eq:Par-SubProblem3}\mathrm{f}),\\
& \sum_{m\in \mathcal{M}} w_{n, m, m} x_{n, m}^2-y_n^{(i-1)}\left(2 y_n-y_n^{(i-1)}\right) \leq \alpha_n, \forall n \in \mathcal{N}, \\
& y_n^2-\sum_{m \in \mathcal{M}} w_{n, m, m} x_{n, m}^{(i-1)}\left(2 x_{n, m}-x_{n, m}^{(i-1)}\right) \leq \beta_n, \forall n \in \mathcal{N}, \\
& x_{n, m}\left(1-2 x_{n, m}^{(i-1)}\right)+\left(x_{n, m}^{(i-1)}\right)^2 \leq \gamma_{n, m}, \forall n \in \mathcal{N}, m \in \mathcal{M},
\end{alignat}
\end{subequations}
where $\rho^{(i-1)}$ is the penalty parameter at the $(i-1)$th iteration. Problem (\ref{eq:Par-SubProblem4}) is a convex problem that can be solved efficiently by an IPT algorithm. The pseudo-code for solving problem (\ref{eq:Par-SubProblem4}) is presented in Algorithm 1. The computational complexity of solving problem (\ref{eq:Par-SubProblem4}) using an IPT algorithm is $\mathcal{O}\left(N^3 M^3\right)$, and the number of iterations of the IPT algorithm is $\mathcal{O}(\sqrt{NM} \log (1/\xi))$, where $\xi$ is the convergence accuracy. Therefore, the total computational complexity of Algorithm 1 is $\mathcal{O}\left((NM)^{3.5} \log (1/\xi)\right)$ \cite{Nesterov39}. Note that the sequence solution $\left\{\mathbf{x}^{(i)}\right\}_{i=1}^{\infty}$ to problem (\ref{eq:Par-SubProblem4}) can converge to a stationary point of problem (\ref{eq:Par-SubProblem2}), as shown in \cite{Lipp40}. Since problem (\ref{eq:Par-SubProblem2}) and problem (\ref{eq:Par-SubProblem0}) are equivalent, Algorithm 1 can also converge to a stationary point of problem (\ref{eq:Par-SubProblem0}).

In summary, the pseudo-code for solving the original problem (9) is provided in Algorithm 2, which is achieved by iteratively solving the resource allocation subproblem and the DNN model partitioning subproblem.

\begin{algorithm}[t]
\caption{PCCP Algorithm for Solving Problem (\ref{eq:Par-SubProblem0})}
\label{alg:1}
\begin{algorithmic}[1]
\State \textbf{Initialize:} Set the initial penalty $\rho^{(0)}>0$, the maximum penalty $\rho_{\max}>0$, the weight $\nu>1$, and the convergence criteria as $\theta_{\text{err}}>0$; Choose an arbitrary initial point $\left\{\mathbf{x}^{(0)}, \mathbf{y}^{(0)}\right\}$ of problem (\ref{eq:Par-SubProblem1}).
\State Set $i=1$.
\Repeat
\State Obtain $\left\{\mathbf{x}^{(i)}, \mathbf{y}^{(i)}\right\}$ by solving the problem (\ref{eq:Par-SubProblem4}) using
\Statex \quad \, an IPT algorithm.
\State Set $\rho^{(i)}=\min \left\{\nu \rho^{(i-1)}, \rho_{\max}\right\}$.
\State Set $i=i+1$.
\Until $\left\|\mathbf{x}^{(i)}-\mathbf{x}^{(i-1)}\right\|<\theta_{\text{err}}$ with $i\geq 1$.
\State Set $ \mathbf{x}=\mathbf{x}^{(i)}$.
\end{algorithmic}
\end{algorithm}

\begin{algorithm}[t]
\caption{Overall Algorithm for Solving Problem (\ref{eq:Problem1})}
\label{alg:2}
\begin{algorithmic}[1]
\State \textbf{Initialize:} Number of mobile devices $N$, partitioning point $M$, communication bandwidth $B$, task deadline $D_n$, risk level $\varepsilon_n$, and the convergence criteria $\theta_{\text{err}}>0$;
\State Set $k=0$.
\State Choose any feasible solution $\left\{\mathbf{x}^{(0)}, \mathbf{b}^{(0)}, \mathbf{f}^{(0)}\right\}$ to problem (\ref{eq:Problem1}).
\Repeat
\State $\emph{Resource allocation subproblem (\ref{eq:RA-SubProblem0}):}$
\State With fixed $\left\{\mathbf{x}^{(k)}\right\}$, problem (\ref{eq:RA-SubProblem0}) is equivalently trans-
\Statex \quad \, formed to problem (\ref{eq:RA-SubProblem1}) by the CCP method.
\State Obtain $\left\{\mathbf{b}^{(k+1)}, \mathbf{f}^{(k+1)}\right\}$ by solving the problem (\ref{eq:RA-SubProblem1})
\Statex \quad \, using an IPT method.
\State $\emph{DNN model partitioning subproblem (\ref{eq:Par-SubProblem0}):}$
\State With fixed $\left\{\mathbf{b}^{(k+1)}, \mathbf{f}^{(k+1)}\right\}$, problem (\ref{eq:Par-SubProblem0}) is equiva-
\Statex \quad \, lently transformed to problem (\ref{eq:Par-SubProblem1}) by the CCP method.
\State Obtain $\left\{\mathbf{x}^{(k+1)}\right\}$ using Algorithm 1.
\State Set $k=k+1$.
\Until The objective value of problem (\ref{eq:Problem1}) meets the conver-
\Statex \quad \quad \, gence criteria $\theta_{\text{err}}$.
\end{algorithmic}
\end{algorithm}

\section{Simulation Results}
In this section, we first give the values of simulation parameters, then show the convergence and complexity of the proposed algorithms, and finally evaluate the performance of the proposed algorithms under different parameter settings.

\subsection{Simulation Setup}
We simulate a $400 \mathrm{~m} \times 400 \mathrm{~m}$ square area with the edge node located at the center of the area. The mobile devices are distributed uniformly and randomly across the coverage area of the edge node. The uplink wireless channel gain between mobile device $n$ and the edge node is modeled as $h_n = 38 + 30 \times \log_{10}r_n$ \cite{3GPP41}, where $h_n$ and $r_n$ are the path-loss (in dB) and distance between device $n$ and the edge node (in meters), respectively. Additionally, the total uplink wireless bandwidth is set to $B = 10$ MHz and $B = 30$ MHz for different DNN models, the transmit power $p_n$ of mobile device $n$ is set to $1$ W, and the noise power density is $N_0 = -174$ dBm/Hz \cite{Shi12}, \cite{Xu15}.

\captionsetup[table]{labelsep=newline}
\begin{table}[!htb]
    \footnotesize
    \caption{\label{tab:II}\textsc{Configurations of DNNs and hardware}}
	\centering
	\begin{tabular}{p{1.5cm}<{\centering}|p{2.8cm}<{\centering}|p{1.6cm}<{\centering}}
		\hline
		\textcolor{black}{\textbf{DNN model}} &\textcolor{black}{\textbf{Mobile device}} &\textcolor{black}{\textbf{VM}} \\
        \hline
        \vspace{-0.06cm} \textcolor{black}{AlexNet} & \textcolor{black}{Jetson Xavier NX CPU $f \in[0.1, 1.2] \mathrm{GHz}$} &\textcolor{black}{GeForce RTX 4080} \\
        \hline
        \vspace{-0.06cm} \textcolor{black}{ResNet152} &\textcolor{black}{Jetson Xavier NX GPU $f \in[0.2, 0.8] \mathrm{GHz}$} &\textcolor{black}{GeForce RTX 4080} \\
        \hline
        \vspace{-0.06cm} \textcolor{black}{ViT-B/32} &\textcolor{black}{Jetson Nano GPU $f \in[0.1, 0.8] \mathrm{GHz}$} &\textcolor{black}{GeForce RTX 4080} \\
        \hline
	\end{tabular}
    \vspace{-0.2cm}
\end{table}

\captionsetup[table]{labelsep=newline}
\begin{table*}[t]
    \scriptsize 
    \caption{\label{tab:III}\textsc{The parameters of ViT-B/32 on Jetson Nano GPU.}}
    \centering
	\begin{tabular}{p{2.2cm}<{\centering}|p{1.0cm}<{\centering} p{1.0cm}<{\centering} p{1.0cm}<{\centering} p{1.0cm}<{\centering} p{1.0cm}<{\centering} p{1.0cm}<{\centering} p{1.0cm}<{\centering} }
		\hline
		\textcolor{black}{Parameter} &\textcolor{black}{point 0} &\textcolor{black}{point 1} &\textcolor{black}{point 2} &\textcolor{black}{point 3} &\textcolor{black}{point 4} &\textcolor{black}{point 5} &\textcolor{black}{point 6} \\
		\hline
		\textcolor{black}{$d_{n, m}$ (MB)} &\textcolor{black}{0.574} &\textcolor{black}{0.146} &\textcolor{black}{0.146} &\textcolor{black}{0.146} &\textcolor{black}{0.146} &\textcolor{black}{0.146} &\textcolor{black}{0.001}\\
		\textcolor{black}{$w_{n, m}$ (GFLOPs)} &\textcolor{black}{--} &\textcolor{black}{3.0954} &\textcolor{black}{3.8114} &\textcolor{black}{5.2435} &\textcolor{black}{7.3916} &\textcolor{black}{8.1077} &\textcolor{black}{8.8253} \\
		\textcolor{black}{$g_{n, m}$ (FLOPs/cycle)} &\textcolor{black}{--} &\textcolor{black}{171.967} &\textcolor{black}{174.837} &\textcolor{black}{175.369} &\textcolor{black}{181.168} &\textcolor{black}{178.191} &\textcolor{black}{135.983} \\
		\textcolor{black}{$v_{n, m}^{\mathrm{loc}}$ $(\mathrm{ms})^2$} &\textcolor{black}{--} &\textcolor{black}{11.059} &\textcolor{black}{18.931} &\textcolor{black}{33.337} &\textcolor{black}{65.814} &\textcolor{black}{75.867} &\textcolor{black}{153.434} \\
        \hline
	\end{tabular}
    \vspace{-0.3cm}
\end{table*}

\captionsetup[table]{labelsep=newline}
\begin{table*}[t]
    \scriptsize 
    \caption{\label{tab:IV}\textsc{The parameters of AlexNet on Jetson Xavier NX CPU.}}
	\centering
	\begin{tabular}{p{2.2cm}<{\centering}|p{1.0cm}<{\centering} p{1.0cm}<{\centering} p{1.0cm}<{\centering} p{1.0cm}<{\centering} p{1.0cm}<{\centering} p{1.0cm}<{\centering} p{1.0cm}<{\centering} p{1.0cm}<{\centering} p{1.0cm}<{\centering}}
		\hline
		\textcolor{black}{Parameter} &\textcolor{black}{point 0} &\textcolor{black}{point 1} &\textcolor{black}{point 2} &\textcolor{black}{point 3} &\textcolor{black}{point 4} &\textcolor{black}{point 5} &\textcolor{black}{point 6} &\textcolor{black}{point 7} &\textcolor{black}{point 8} \\
		\hline
		\textcolor{black}{$d_{n, m}$ (MB)} &\textcolor{black}{0.574} &\textcolor{black}{0.74} &\textcolor{black}{0.18} &\textcolor{black}{0.53} &\textcolor{black}{0.12} &\textcolor{black}{0.25} &\textcolor{black}{0.17} &\textcolor{black}{0.04} &\textcolor{black}{0.001} \\
		\textcolor{black}{$w_{n, m}$ (GFLOPs)} &\textcolor{black}{--} &\textcolor{black}{0.1407} &\textcolor{black}{0.1411} &\textcolor{black}{0.5891} &\textcolor{black}{0.5894} &\textcolor{black}{0.8137} &\textcolor{black}{1.3122} &\textcolor{black}{1.3123} &\textcolor{black}{1.4214} \\
		\textcolor{black}{$g_{n, m}$ (FLOPs/cycle)} &\textcolor{black}{--} &\textcolor{black}{6.8994} &\textcolor{black}{6.3283} &\textcolor{black}{13.6064} &\textcolor{black}{13.1861} &\textcolor{black}{14.6624} &\textcolor{black}{16.4237} &\textcolor{black}{16.1219} &\textcolor{black}{7.1037} \\
		\textcolor{black}{$v_{n, m}^{\mathrm{loc}}$ $(\mathrm{ms})^2$} &\textcolor{black}{--} &\textcolor{black}{37.341} &\textcolor{black}{43.084} &\textcolor{black}{59.616} &\textcolor{black}{63.942} &\textcolor{black}{74.801} &\textcolor{black}{95.073} &\textcolor{black}{98.876} &\textcolor{black}{105.886} \\
        \hline
	\end{tabular}
    \vspace{-0.2cm}
\end{table*}

\captionsetup[table]{labelsep=newline}
\begin{table*}[t]
    \scriptsize 
    \caption{\label{tab:V}\textsc{The parameters of ResNet152 on Jetson Xavier NX GPU.}}
    \centering
	\begin{tabular}{p{2.2cm}<{\centering}|p{1.0cm}<{\centering} p{1.0cm}<{\centering} p{1.0cm}<{\centering} p{1.0cm}<{\centering} p{1.0cm}<{\centering} p{1.0cm}<{\centering} p{1.0cm}<{\centering} p{1.0cm}<{\centering} p{1.0cm}<{\centering} p{1.0cm}<{\centering}}
		\hline
		\textcolor{black}{Parameter} &\textcolor{black}{point 0} &\textcolor{black}{point 1} &\textcolor{black}{point 2} &\textcolor{black}{point 3} &\textcolor{black}{point 4} &\textcolor{black}{point 5} &\textcolor{black}{point 6} &\textcolor{black}{point 7} &\textcolor{black}{point 8} &\textcolor{black}{point 9}\\
		\hline
		\textcolor{black}{$d_{n, m}$ (MB)} &\textcolor{black}{0.574} &\textcolor{black}{3.06} &\textcolor{black}{0.77} &\textcolor{black}{1.53} &\textcolor{black}{0.38} &\textcolor{black}{0.19} &\textcolor{black}{0.19} &\textcolor{black}{0.19} &\textcolor{black}{0.1} &\textcolor{black}{0.001}\\
		\textcolor{black}{$w_{n, m}$ (GFLOPs)} &\textcolor{black}{--} &\textcolor{black}{0.2392} &\textcolor{black}{1.4864} &\textcolor{black}{3.6585} &\textcolor{black}{5.3099} &\textcolor{black}{9.9984} &\textcolor{black}{13.9389} &\textcolor{black}{17.8794} &\textcolor{black}{21.9228} &\textcolor{black}{23.1064} \\
		\textcolor{black}{$g_{n, m}$ (FLOPs/cycle)} &\textcolor{black}{--} &\textcolor{black}{315.4525} &\textcolor{black}{309.6695} &\textcolor{black}{323.7640} &\textcolor{black}{329.8090} &\textcolor{black}{325.6815} &\textcolor{black}{324.1615} &\textcolor{black}{322.7340} &\textcolor{black}{318.6457} &\textcolor{black}{307.6753} \\
		\textcolor{black}{$v_{n, m}^{\mathrm{loc}}$ $(\mathrm{ms})^2$} &\textcolor{black}{--} &\textcolor{black}{0.097} &\textcolor{black}{1.310} &\textcolor{black}{5.677} &\textcolor{black}{13.934} &\textcolor{black}{14.076} &\textcolor{black}{15.881} &\textcolor{black}{23.408} &\textcolor{black}{32.256} &\textcolor{black}{32.727} \\
        \hline
	\end{tabular}
    \vspace{-0.3cm}
\end{table*}

Three widely-used DNNs, AlexNet \cite{Krizhevsky23}, ResNet152 \cite{He24} and ViT-B/32 \cite{Dosovitskiy50}, are considered. The three DNNs are fully deployed on mobile devices and the MEC server. The task of the mobile device is image recognition, which is extracted from the object recognition dataset CIFAR-10 \cite{Krizhevsky42}. The processing unit of mobile devices adopts Jetson Xavier NX CPU and GPU \cite{Xavier43}, as well as Jetson Nano GPU \cite{Nano51}. We assume that AlexNet is deployed on the Jetson Xavier NX CPU, ResNet152 is deployed on Jetson Xavier NX GPU, and ViT-B/32 is deployed on Jetson Nano GPU. The VM assigned to the mobile device uses the GeForce RTX 4080. Specific configurations are shown in Table II.

The energy efficiency coefficient $\kappa_n$ of Jetson Xavier NX CPU and GPU is evaluated using the power testing tool Tegrastats of NVIDIA \cite{Tegrastats44}. Specifically, Jetson Xavier NX is first set to a fixed power consumption mode. Then, the power consumption of the CPU and GPU at different frequencies is measured, and finally, $\kappa_n$ is obtained based on the measured data. By estimation, the average $\kappa_n$ of the Jetson Xavier NX CPU, Jetson Xavier NX GPU and Jetson Nano GPU are $0.8 \times 10^{-27} \mathrm{W}/(\text{cycle}/\mathrm{sec})^3$, $2.8 \times 10^{-27} \mathrm{W}/(\text{cycle}/\mathrm{sec})^3$, and $3.2 \times 10^{-27} \mathrm{W}/(\text{cycle}/\mathrm{sec})^3$, respectively.

AlexNet, ResNet152 and ViT-B/32 are divided into 8, 9 and 6 blocks, corresponding to 9, 10 and 7 partitioning points, respectively. The feature data size of each block can be calculated based on its output data shape. The mean inference time for each block is obtained through 500 experiments, and then the variance and covariance can also be calculated based on the mean and measured data. The specific parameters are shown in Table III, IV and V. Unless otherwise specified, the above parameters are used by default.

To evaluate the performance of Algorithm 1, we consider the following two policies as benchmark:
\begin{enumerate}
  \item {Random policy:} the DNN partitioning point is randomly selected from the set of points where the feature data is smaller than the raw data.
  \item {Optimal policy:} the DNN partitioning point is obtained using the exhaustive search method, which can find the optimal partitioning point, but its computational complexity is exponential.
\end{enumerate}

To evaluate the performance of Algorithm 2, we consider the following two methods as benchmark:
\begin{enumerate}
  \item {Worst-case optimization:} the upper bound of $t_{n, m}^{\mathrm{loc}}$ and $t_{n, m}^{\mathrm{vm}}$ obtained by the experiment is taken as the inference time, and the task deadline is not allowed to be violated.
  \item {Mean-value optimization:} the inference time only adopts the mean value, without considering the uncertainty in inference time. This is a primitive method that bypasses the uncertainty, without any probabilistic guarantees.
\end{enumerate}

\subsection{Convergence and Complexity}
\begin{figure}[t]
  \captionsetup{font={small}}
  \centering
  \includegraphics[width=0.35\textwidth]{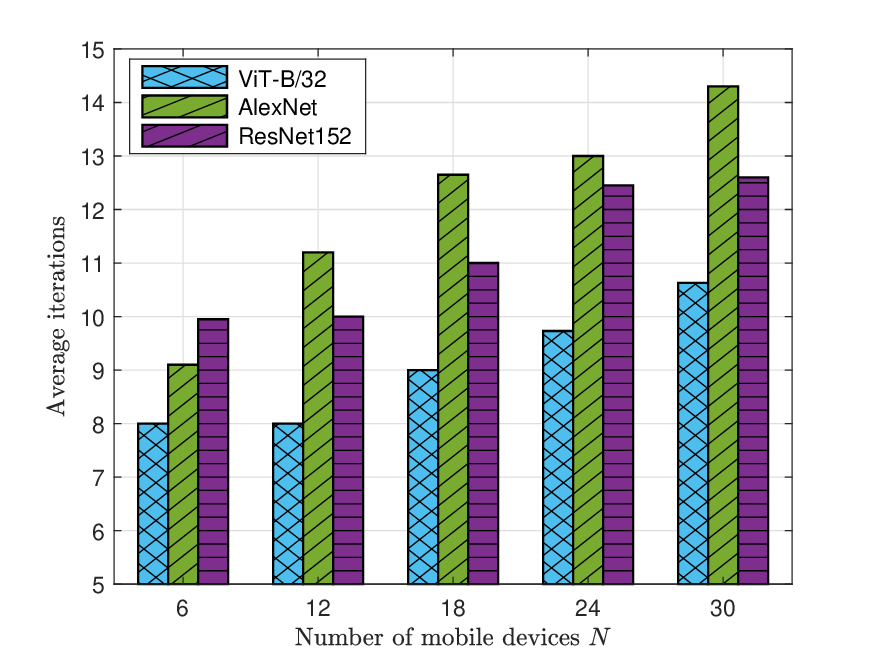}\\
  \caption{The average number of iterations of Algorithm 1 under different numbers of mobile devices in the three models with ViT-B/32, AlexNet, and ResNet152.}
  \label{fig:ConvAlg1}
  \vspace{-0.3cm}
\end{figure}

First, we show the convergence of the proposed algorithms. Fig. \ref{fig:ConvAlg1} illustrates the average number of iterations of Algorithm 1 versus the number of mobile devices. Although the number of iterations of Algorithm 1 cannot be analytically characterized, we can see from Fig. \ref{fig:ConvAlg1} that even when the number of devices $N=30$, Algorithm 1 can terminate after a few iterations. Moreover, the average number of iterations for ViT-B/32, AlexNet, and ResNet152 are not significantly different. In addition, the average number of iterations of Algorithm 1 increases slightly as the number of mobile devices increases significantly. This indicates that Algorithm 1 based on PCCP has better scalability.

\begin{figure}[t]
  \captionsetup{font={small}}
  \centering
  \includegraphics[width=0.48\textwidth]{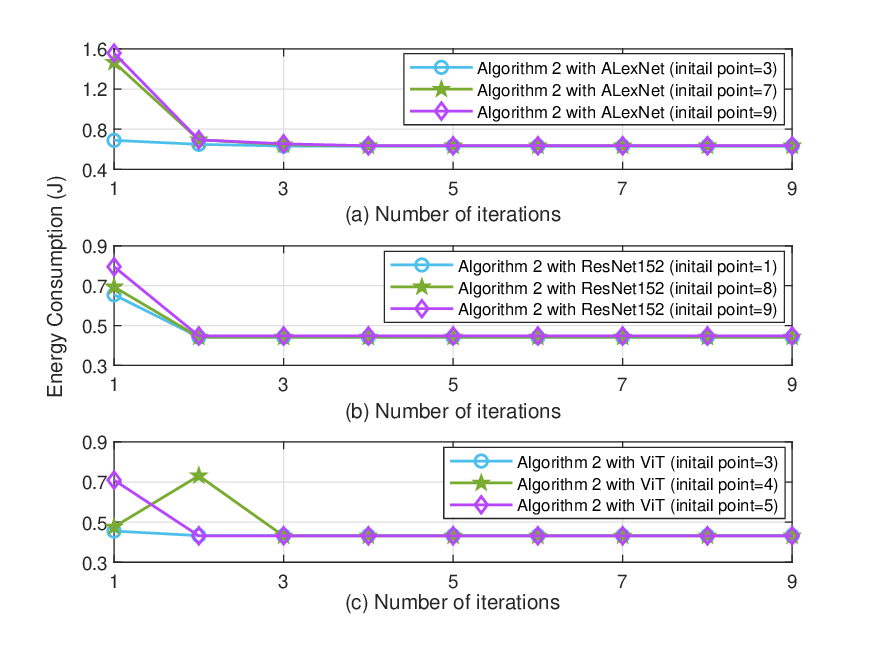}\\
  \caption{The convergence trajectories of Algorithm 2 for AlexNet ($D_n=220$ ms), ResNet152 ($D_n=160$ ms), and ViT-B/32 ($D_n=130$ ms).}
  \label{fig:ConvAlg2}
  \vspace{-0.3cm}
\end{figure}

Fig. \ref{fig:ConvAlg2}(a) and Fig. \ref{fig:ConvAlg2}(b) illustrate the convergence trajectories of Algorithm 2 from different initial points in AlexNet, ResNet152, and ViT-B/32 models, respectively. We select three different points as the initial points from all the partitioning points of AlexNet, ResNet152, and ViT-B/32, respectively. For example, the initial points for AlexNet are 3, 7, and 9; for ResNet152, they are 1, 8, and 9; and for ViT-B/32, they are 3, 4, and 5. From Fig. \ref{fig:ConvAlg2}, it can be observed that the advantage of using Algorithm 2 is its ability to converge quickly in the early stages of iteration. In addition, Algorithm 2 almost converges to the same objective function value for different initial points.

\begin{figure}[t]
  \captionsetup{font={small}}
  \centering
  \includegraphics[width=0.35\textwidth]{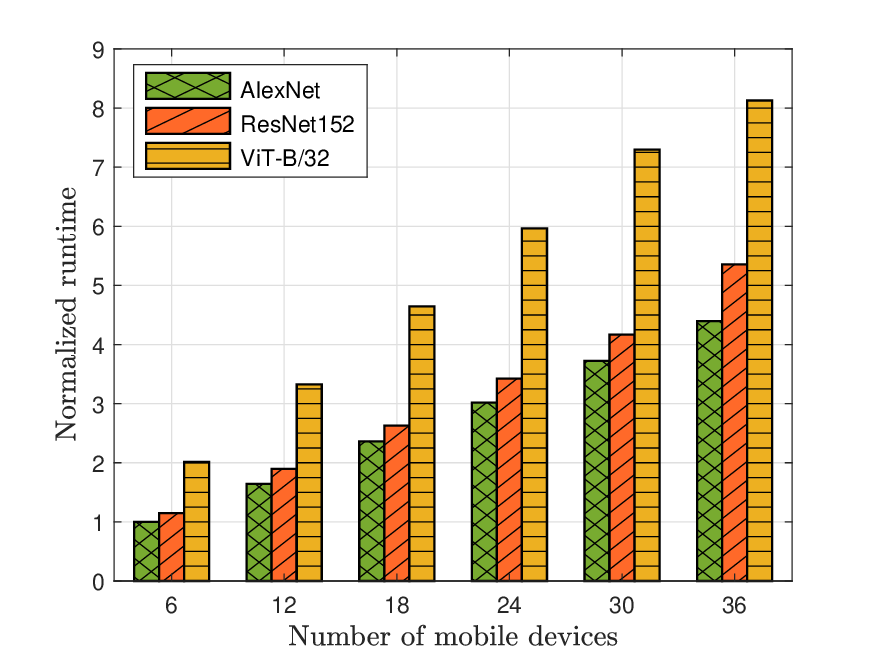}\\
  \caption{The average runtime of Algorithm 2 under different number of mobile devices.}
  \label{fig:Runtime}
  \vspace{-0.3cm}
\end{figure}

Then, we show the computational complexity of the proposed algorithms. Fig. \ref{fig:Runtime} illustrates the runtime of Algorithm 2 on AlexNet, ResNet152, and ViT-B/32. The simulation experiments are implemented using MATLAB and conducted on a laptop computer with an Intel Core i7-8700 3.2 GHz CPU and 16 GB RAM. In Fig. \ref{fig:Runtime}, the runtime for AlexNet with $N = 6$ is normalized to 1. From Fig. \ref{fig:Runtime}, it can be seen that the runtime of the proposed algorithm increases linearly with the number of mobile devices, despite the exponentially growing search space. Since the ResNet152 model has 10 partitioning points, its average runtime is slightly higher than that of the AlexNet model, which has 9 partitioning points. Despite having only 7 partitioning points, the ViT-B/32 model actually exhibits a higher runtime. It is because, as shown in Fig. \ref{fig:ConvAlg2}, the ViT-B/32 model requires more iterations than both AlexNet and ResNet152. Combining this with the computational complexity analysis in Section IV, we observed that the complexity of the proposed Algorithm 1 and Algorithm 2 are polynomial time with respect to the number of partitioning points and mobile devices.

In addition, we found that although the theoretical computational complexity of the IPT method is proportional to $M$ and $N$ to the 3.5th power, i.e., $\mathcal{O}\left((N M)^{3.5} \log (1 / \xi)\right)$, Fig. \ref{fig:Runtime} shows a much lower practical increase in runtime. Specifically, when the number of devices increases from $N = 6$ to $N = 36$, the runtimes of AlexNet, ResNet152, and ViT-B/32 increase by less than 5 times. This observed runtime growth is much lower than what would be expected based solely on the theoretical computational complexity. It may be related to factors such as the optimization methods employed by the solver and the performance of the hardware platform. For larger-scale networks, such as $M \gg 10$, and $N \gg 36$, we can consider combining the proposed algorithm with distributed optimization to reduce the computational complexity of the algorithm.

\subsection{Performance Evaluation}
In this subsection, we investigate the impact of variance approximation, different device numbers, risk levels, and task deadlines on the total energy consumption under the AlexNet, ViT-B/32, and ResNet152 models. In addition, we further analyze the violation probability of the task deadline under varying risk levels.

\subsubsection{Impact of variance approximation}
\begin{figure}[t]
    \centering
    \begin{subfigure}[b]{0.24\textwidth}
        \centering
        \includegraphics[width=\textwidth]{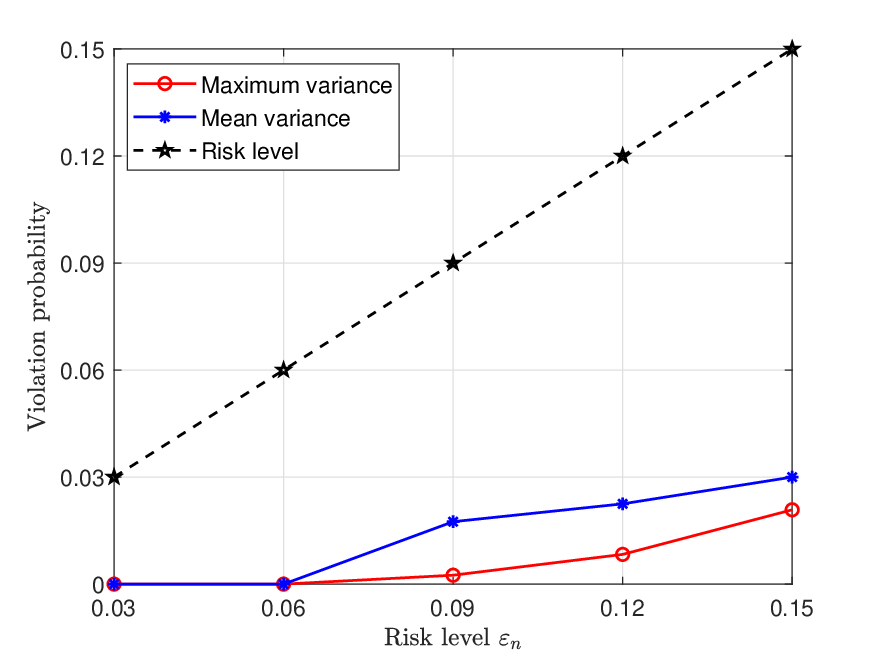}
        \caption{\scriptsize{Violation probability with AlexNet}}
        \label{fig:sub1}
    \end{subfigure}
    \hfill
    \begin{subfigure}[b]{0.24\textwidth}
        \centering
        \includegraphics[width=\textwidth]{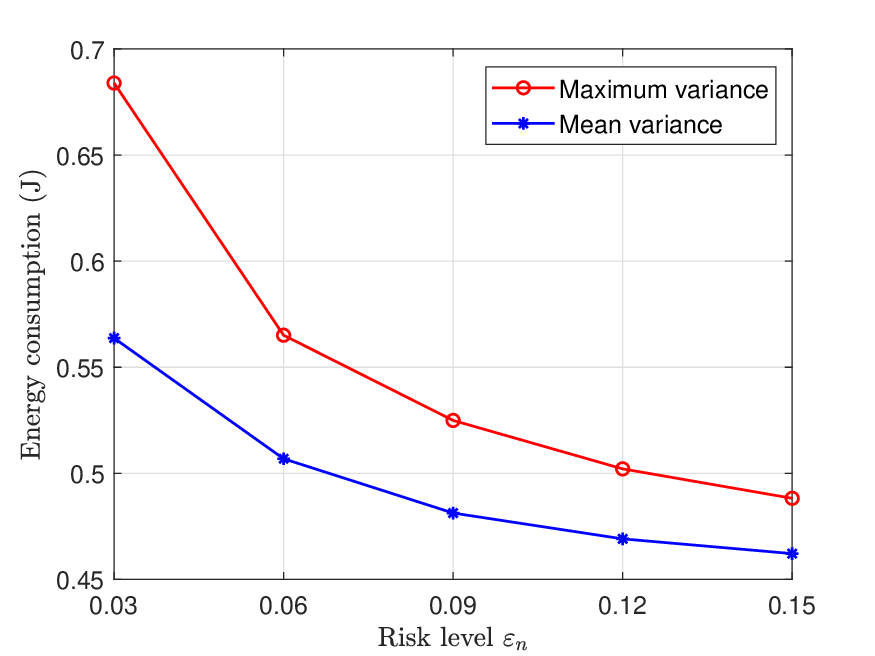}
        \caption{\scriptsize{Energy consumption with AlexNet}}
        \label{fig:sub2}
    \end{subfigure}

    \begin{subfigure}[b]{0.24\textwidth}
        \centering
        \includegraphics[width=\textwidth]{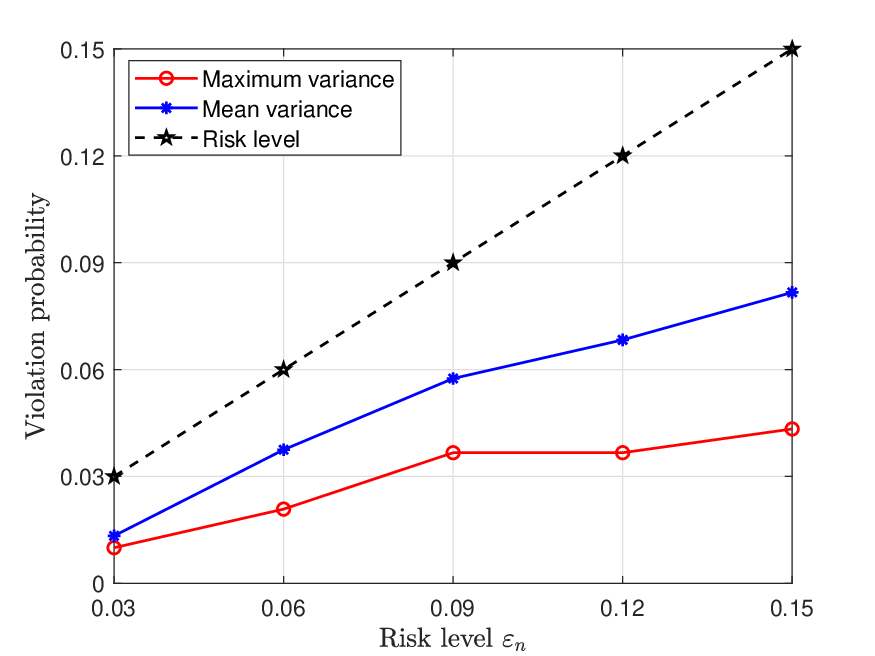}
        \caption{\scriptsize{Violation probability with ViT-B/32}}
        \label{fig:sub3}
    \end{subfigure}
    \hfill
    \begin{subfigure}[b]{0.24\textwidth}
        \centering
        \includegraphics[width=\textwidth]{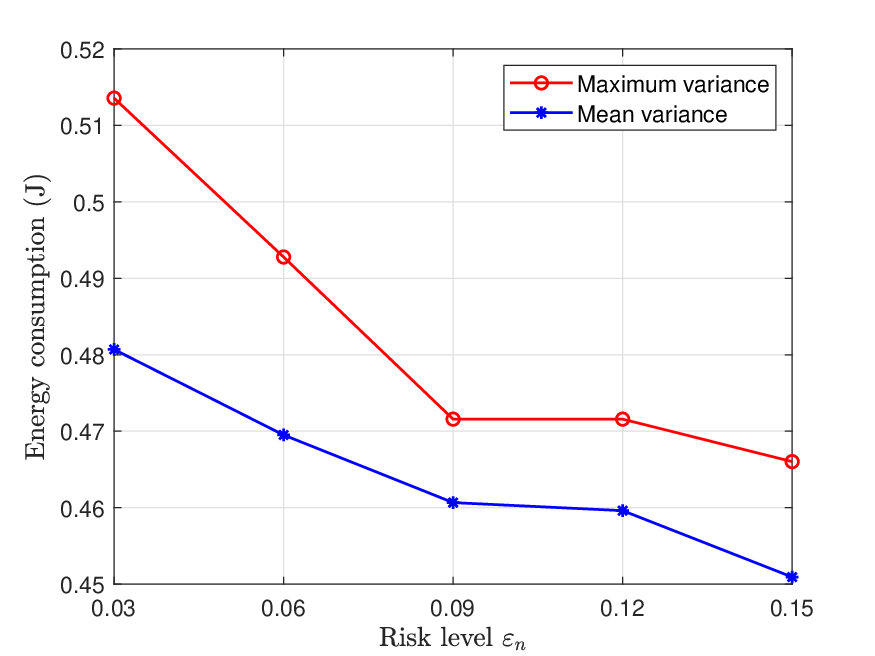}
        \caption{\scriptsize{Energy consumption with ViT-B/32}}
        \label{fig:sub4}
    \end{subfigure}

    \begin{subfigure}[b]{0.24\textwidth}
        \centering
        \includegraphics[width=\textwidth]{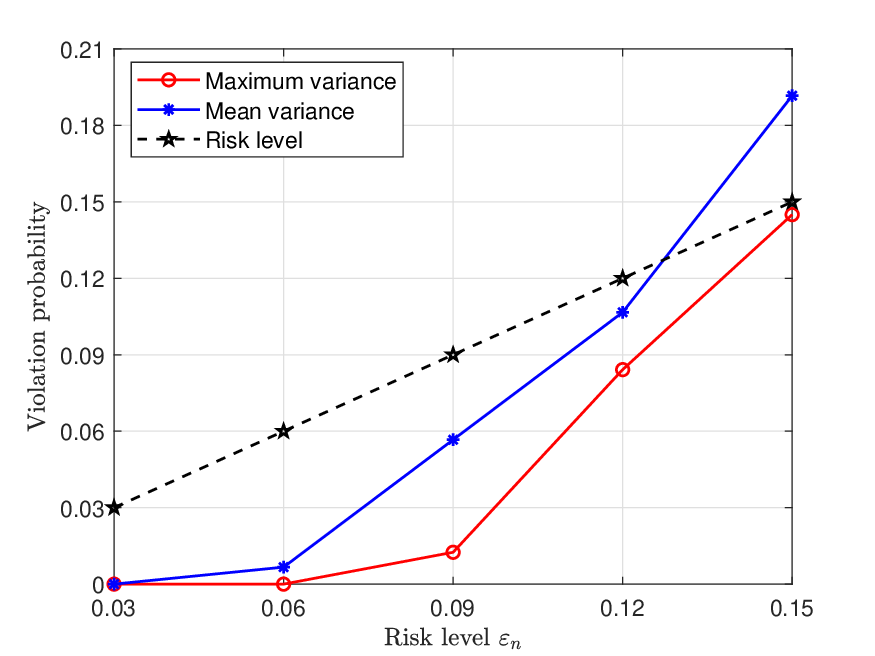}
        \caption{\scriptsize{Violation probability with ResNet152}}
        \label{fig:sub5}
    \end{subfigure}
    \hfill
    \begin{subfigure}[b]{0.24\textwidth}
        \centering
        \includegraphics[width=\textwidth]{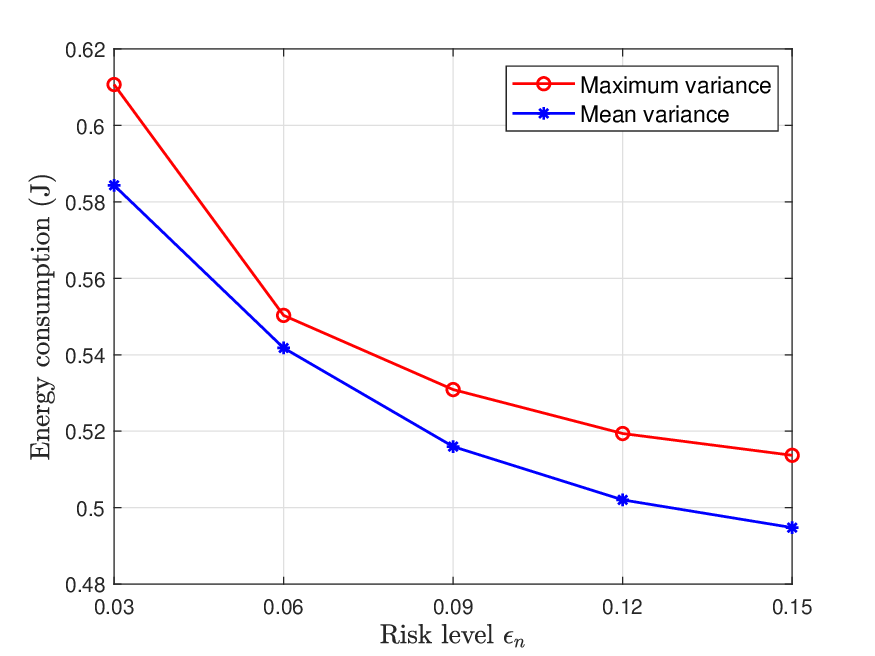}
        \caption{\scriptsize{Energy consumption with ResNet152}}
        \label{fig:sub6}
    \end{subfigure}
    \caption{Impact of variance approximation on AlexNet ($N = 12$, $B = 10$ MHz, $D_n = 180$ ms), ViT-B/32 ($N = 12$, $B = 20$ MHz, $D_n = 100$ ms), and ResNet152 ($N = 12$, $B = 30$ MHz, $D_n = 115$ ms).}
    \label{fig:ApproxVar}
    \vspace{-0.3cm}
\end{figure}

The approximation performance of (\ref{eq:VarFit}) and (\ref{eq:CoVarFit}) in Section IV. B is evaluated below. Fig. \ref{fig:ApproxVar} shows the energy consumption and violation probability of AlexNet, ResNet152, and ViT-B/32 using the maximum and mean variance. The mean variance method is an approximation of (\ref{eq:VarFit}) and (\ref{eq:CoVarFit}). The maximum variance method is approximated as follows:
\begin{equation*}
\left\{\begin{array}{l}
v_{n, m}^{\text {loc }}=\max _{\forall f_n \in \mathcal{F}}\left\{v_{n, m}^{\text {loc }}\left(f_n\right)\right\}, \\
w_{n, m, m^{\prime}}=\max_{\forall f_n \in \mathcal{F}}\left\{w_{n, m, m^{\prime}}\left(f_n\right)\right\}.
\end{array}\right.
\end{equation*}
where $\forall n \in \mathcal{N}, \forall m, m^{\prime} \in \mathcal{M}$.

In Figs. \ref{fig:ApproxVar} (a), (c), and (e), we first show the violation probability as $\varepsilon_n$ increases. It can be observed that the maximum variance method has a lower violation probability than the mean variance method, as the maximum variance method is more conservative. For AlexNet and ResNet152, even when $\varepsilon_n$ increases to 0.15, the violation probability of the mean variance method remains lower than the risk level. However, for ViT-B/32, when $\varepsilon_n$ exceeds 0.13, the violation probability of the mean variance method is slightly higher than the risk level. Then, in Figs. \ref{fig:ApproxVar} (b), (d), and (f), we illustrate the changes in energy consumption as $\varepsilon_n$ increases. It can be observed that the mean variance method results in lower energy consumption compared to the maximum variance method. For AlexNet, ResNet, and ViT-B/32, the mean variance method can reduce energy consumption by up to $17\%$, $6.8\%$, and $4\%$, respectively, compared to the maximum variance method when $\varepsilon_n = 0.03$.

\subsubsection{Performance of Algorithm 1}

\begin{figure}[htbp]
    \centering
    \begin{subfigure}[b]{0.24\textwidth}
        \centering
        \includegraphics[width=\textwidth]{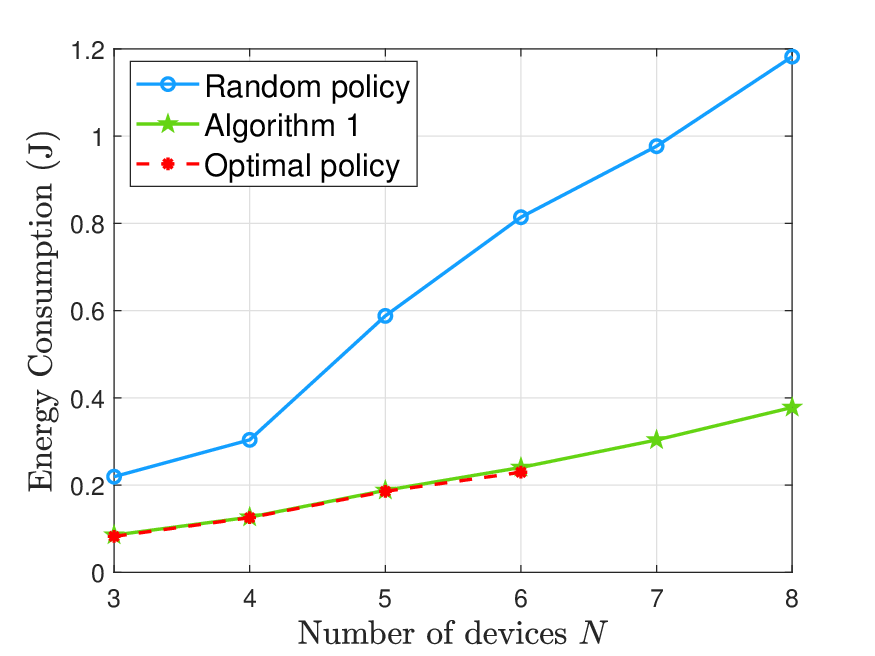}
        \caption{\scriptsize{AlexNet}}
        \label{fig:a}
    \end{subfigure}
    \hfill
    \begin{subfigure}[b]{0.24\textwidth}
        \centering
        \includegraphics[width=\textwidth]{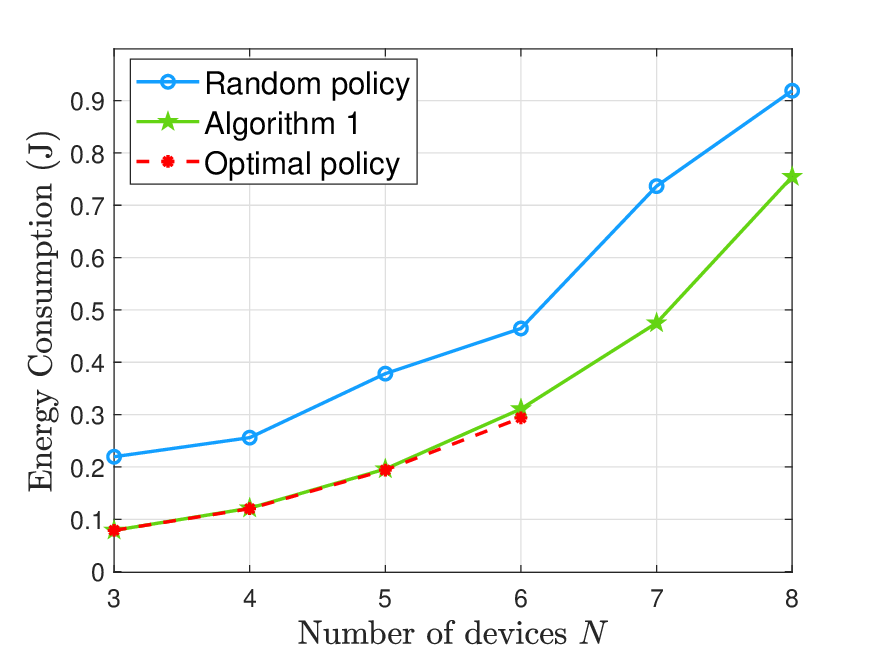}
        \caption{\scriptsize{ResNet152}}
        \label{fig:b}
    \end{subfigure}
    \hfill
    \begin{subfigure}[b]{0.24\textwidth}
        \centering
        \includegraphics[width=\textwidth]{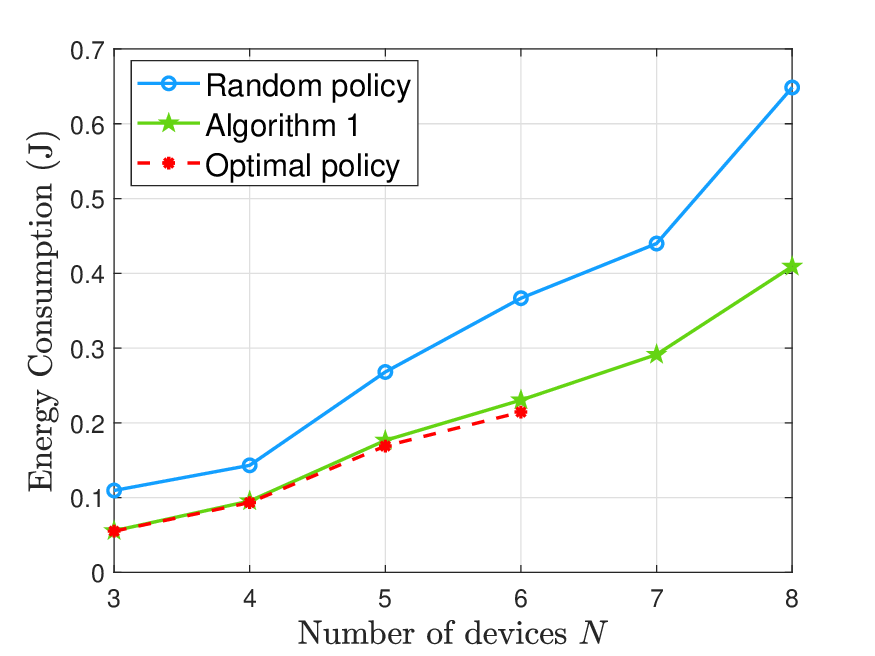}
        \caption{\scriptsize{ViT-B/32}}
        \label{fig:c}
    \end{subfigure}
    \caption{Performance of Algorithm 1 for AlexNet ($D_n=200$ ms, $B=5$ MHz), ResNet152 ($D_n=150$ ms, $B=15$ MHz), and ViT-B/32 ($D_n=130$ ms, $B=10$ MHz).}
    \label{fig:Alg1ES}
    \vspace{-0.3cm}
\end{figure}

Fig. \ref{fig:Alg1ES} evaluates the performance of Algorithm 1 under different DNN models. Firstly, we can observe that the total energy consumption increases with the number of mobile devices. Compared to the random policy, the total energy consumption of Algorithm 1 is much lower. Moreover, as the number of devices increases, the energy consumption of Algorithm 1 increases at a slower rate than the random policy. Secondly, the performance of the proposed Algorithm 1 is very close to the optimal policy. The computational complexity of the optimal policy is $\mathcal{O}\left(M^N\right)$, which is exponential, while the proposed PCCP-based Algorithm 1 can find a stationary point of the DNN model partitioning subproblem, and its computational complexity is polynomial.

\subsubsection{Impact of risk levels}
\begin{figure*}[!htb]
    \captionsetup{font={small}}
    \centering
    \begin{subfigure}[b]{0.32\textwidth}
        \includegraphics[width=1.0\textwidth]{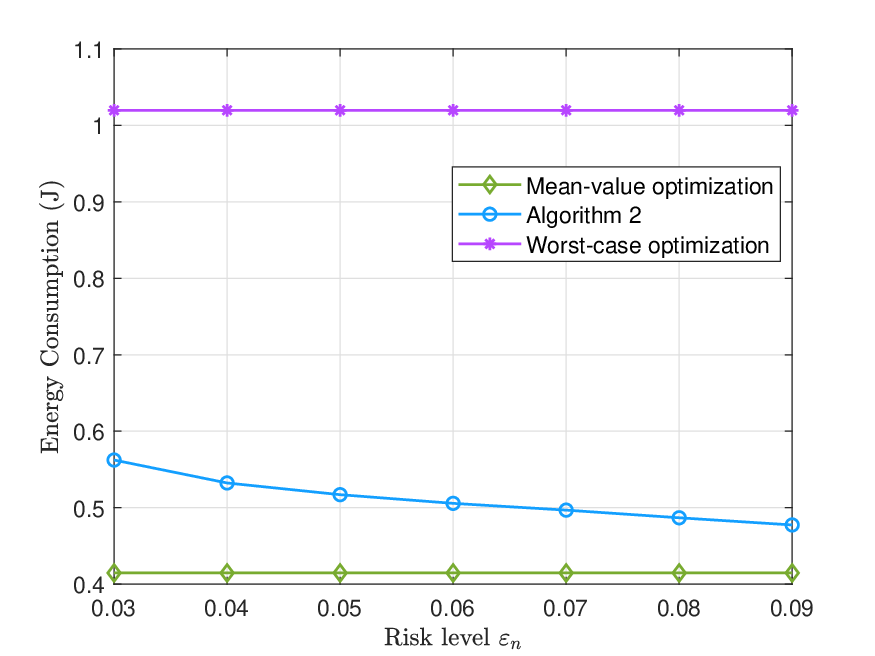}
        \caption{}
        \label{fig:AlexEpsilon}
    \end{subfigure}
    \begin{subfigure}[b]{0.32\textwidth}
        \includegraphics[width=1.0\textwidth]{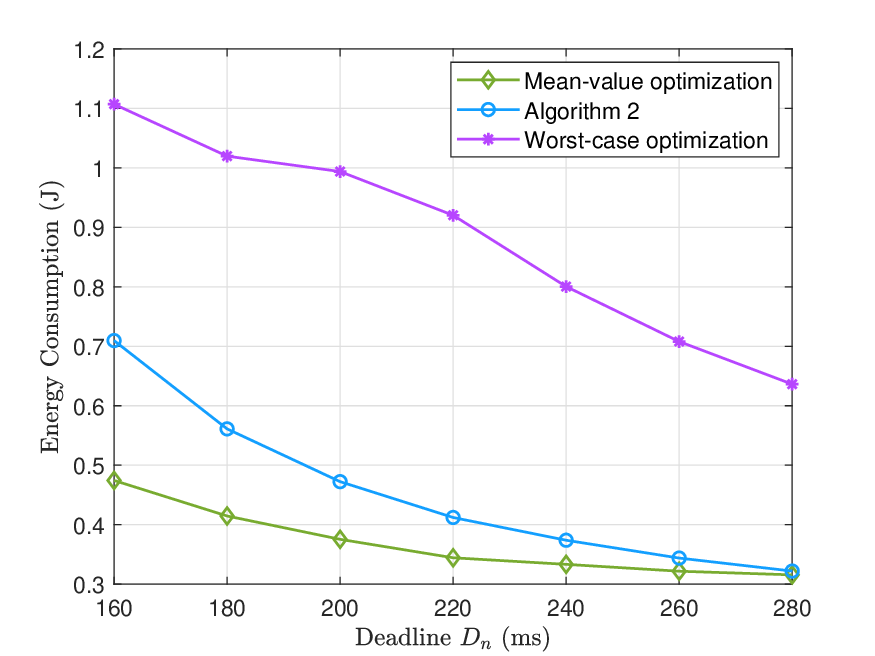}
        \caption{}
        \label{fig:AlexDeadline}
    \end{subfigure}
    \begin{subfigure}[b]{0.32\textwidth}
        \includegraphics[width=1.0\textwidth]{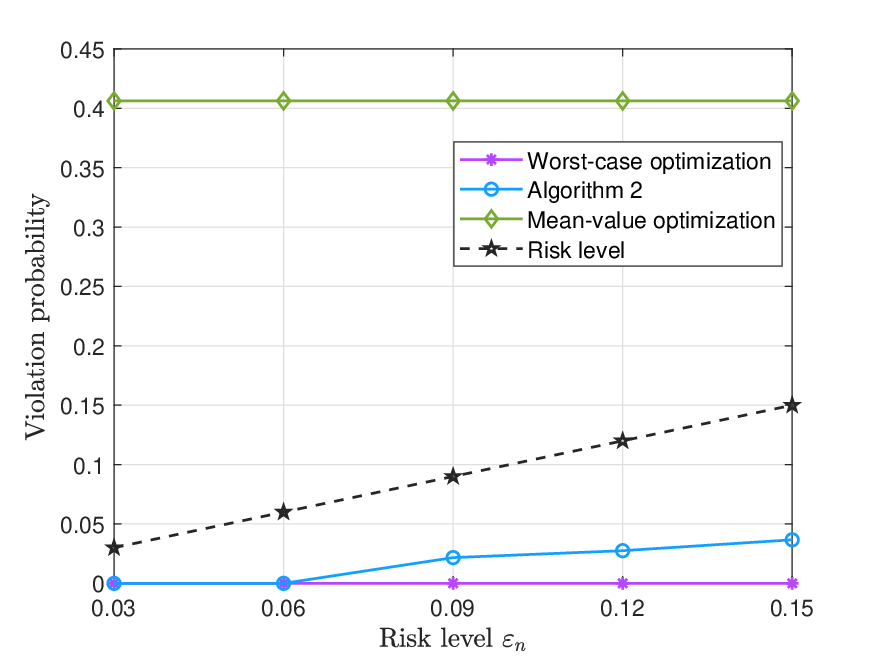}
        \caption{}
        \label{fig:AlexViolation}
    \end{subfigure}
    \caption{The performance of the proposed policy on AlexNet ($N=12$ and $B=10$ MHz). (a) Energy consumption under different risk levels with $D_n=180$ ms. (b) Energy consumption under different deadlines with $\varepsilon_n=0.03$. (c) Deadline violation probability under different risk levels with $D_n=180$ ms.}
    \label{fig:AlexNet}
\end{figure*}

\begin{figure*}[!htb]
    \captionsetup{font={small}}
    \centering
    \begin{subfigure}[b]{0.32\textwidth}
        \includegraphics[width=1.0\textwidth]{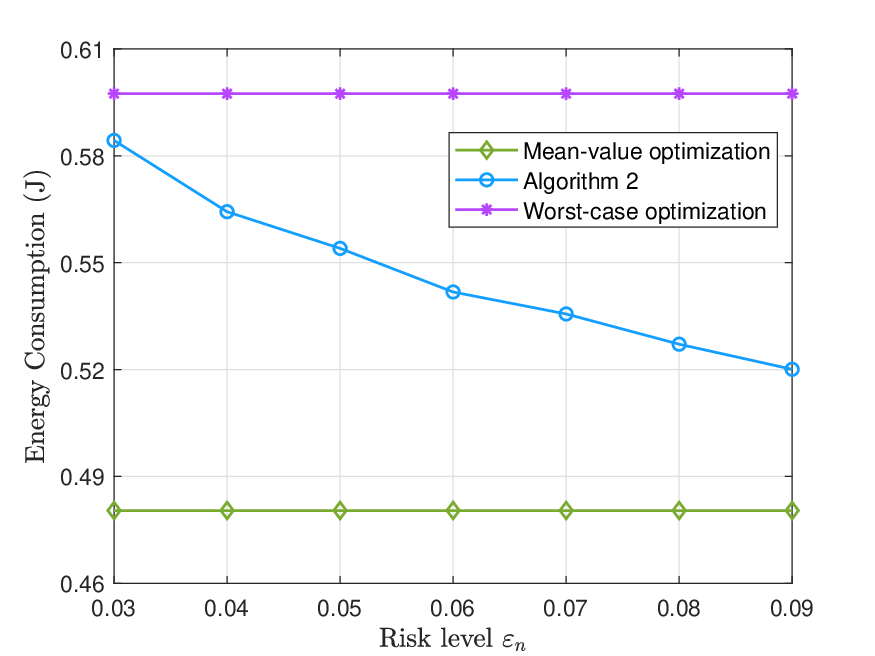}
        \caption{}
        \label{fig:ResNetEpsilon}
    \end{subfigure}
    \begin{subfigure}[b]{0.32\textwidth}
        \includegraphics[width=1.0\textwidth]{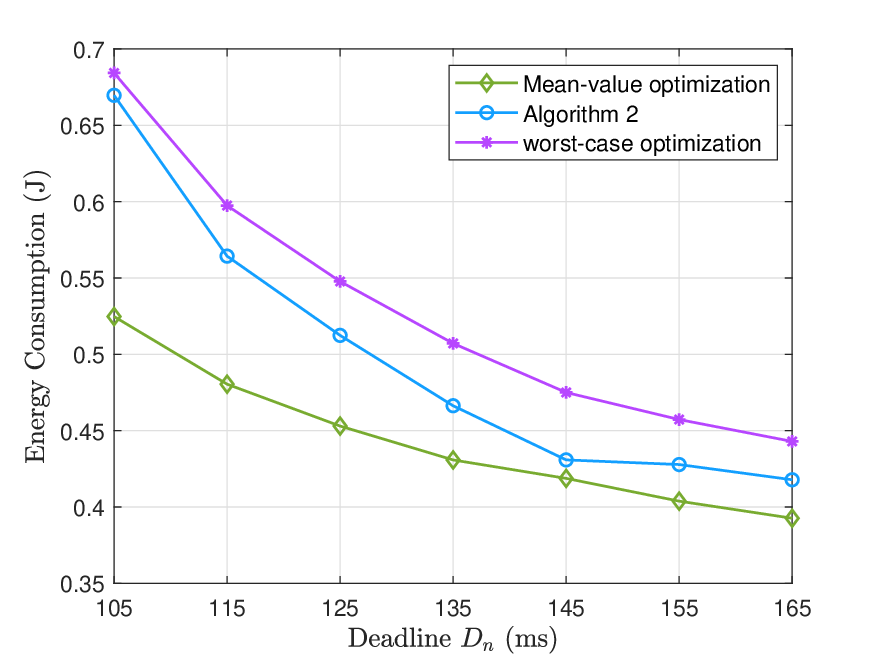}
        \caption{}
        \label{fig:ResNetDeadline}
    \end{subfigure}
    \begin{subfigure}[b]{0.32\textwidth}
        \includegraphics[width=1.0\textwidth]{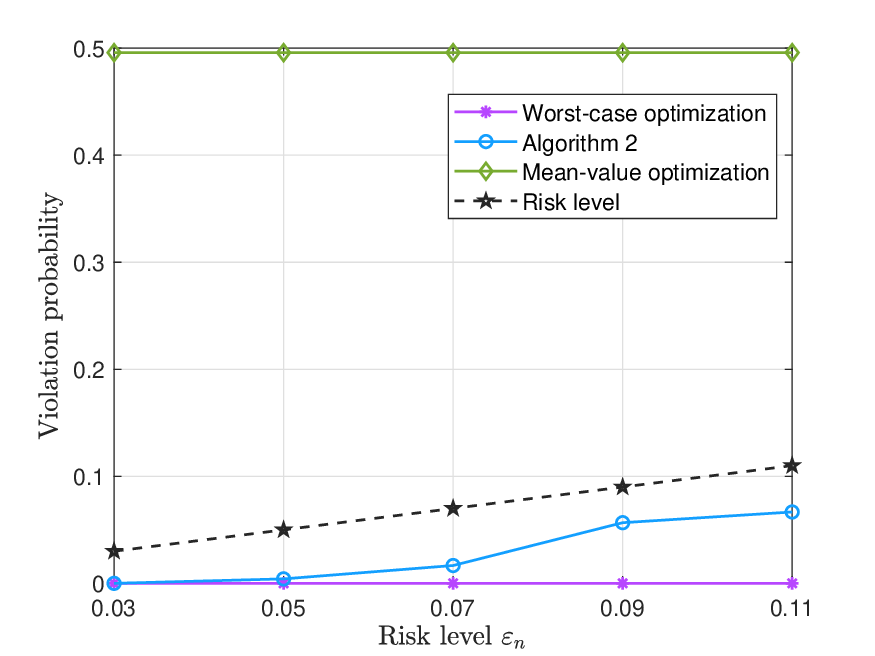}
        \caption{}
        \label{fig:ResNetViolation}
    \end{subfigure}
    \caption{The performance of the proposed policy on ResNet152 ($N=12$ and $B=30$ MHz). (a) Energy consumption under different risk levels with $D_n=115$ ms. (b) Energy consumption under different deadlines with $\varepsilon_n=0.04$. (c) Deadline violation probability under different risk levels with $D_n=115$ ms.}
    \label{fig:ResNet152}
\end{figure*}

\begin{figure*}[!htb]
    \captionsetup{font={small}}
    \centering
    \begin{subfigure}[b]{0.32\textwidth}
        \includegraphics[width=1.0\textwidth]{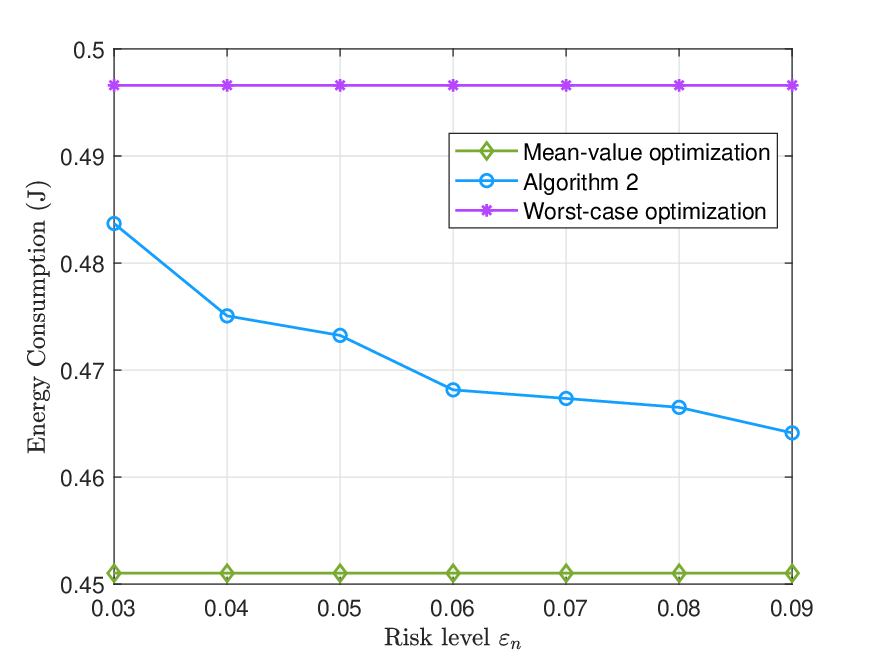}
        \caption{}
        \label{fig:ViTEpsilon}
    \end{subfigure}
    \begin{subfigure}[b]{0.32\textwidth}
        \includegraphics[width=1.0\textwidth]{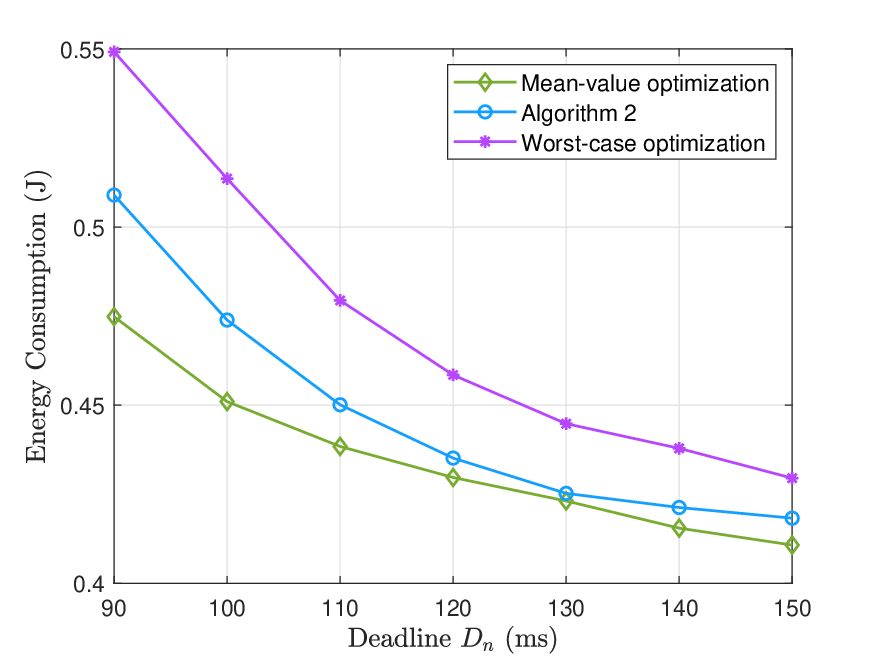}
        \caption{}
        \label{fig:ViTDeadline}
    \end{subfigure}
    \begin{subfigure}[b]{0.32\textwidth}
        \includegraphics[width=1.0\textwidth]{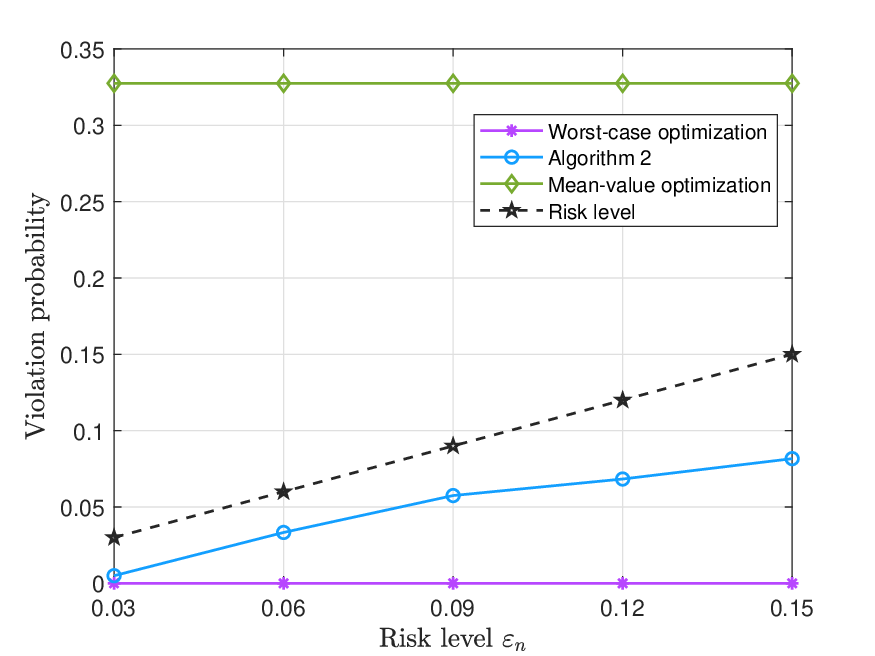}
        \caption{}
        \label{fig:ViTViolation}
    \end{subfigure}
    \caption{The performance of the proposed policy on ViT-B/32 ($N=12$ and $B=20$ MHz). (a) Energy consumption under different risk levels with $D_n=100$ ms. (b) Energy consumption under different deadlines with $\varepsilon_n=0.05$. (c) Deadline violation probability under different risk levels with $D_n=100$ ms.}
    \label{fig:ViT-B/32}
\end{figure*}

Fig. \ref{fig:AlexNet}(a), Fig. \ref{fig:ResNet152}(a), and Fig. \ref{fig:ViT-B/32}(a) show the total energy consumption at different risk levels. Currently, there is no effective solution that guarantees the deadline for DNN partitioning under inference time uncertainty. For comparison, we compare the worst-case and mean optimization methods with our proposed Algorithm 2.

As shown in Fig. \ref{fig:AlexNet}(a), the total energy consumption of Algorithm 2 is always lower than that of the worst-case optimization. Even when the risk level $\varepsilon=0.03$, total energy consumption can be reduced by nearly $44.8\%$. When the risk level increases to 0.09, the total energy consumption of Algorithm 2 saves $53.1\%$ compared to the worst-case optimization. As the risk level increases from 0.03 to 0.09, we observe that the total energy consumption monotonically decreases, which is as expected with (\ref{eq:RA-Determi}) and (\ref{eq:Par-Determi}) that we derived. From (\ref{eq:RA-Determi}), it can be observed that under a given partitioning decision, $\sigma_n$ decreases as $\varepsilon_n$ increases, which means that the variance term of the uncertainty in inference time (i.e., the second term on the left side of (\ref{eq:RA-Determi})) becomes smaller. Mobile devices can save energy consumption by reducing CPU/GPU frequency. Similarly, it can be seen from (\ref{eq:Par-Determi}) that under given communication and computing resources, mobile devices can save energy consumption by selecting appropriate DNN partitioning points.

Fig. \ref{fig:ResNet152}(a) and Fig. \ref{fig:ViT-B/32}(a) show that the energy consumption of ResNet152 and ViT-B/32 also gradually decreases as $\varepsilon_n$ increases. At $\varepsilon_n = 0.06$, compared to the worst-case optimization, ResNet152 and ViT-B/32 can save energy consumption by nearly $9.3\%$ and $5.8\%$, respectively. As shown in Figs. \ref{fig:AlexNet}(c), \ref{fig:ResNet152}(c), and \ref{fig:ViT-B/32}(c), similar to the worst-case optimization, the violation probability of AlexNet and ResNet152 remains close to 0 even when $\varepsilon_n = 0.06$, while the violation probability of ViT-B/32 is only $3.5\%$. Moreover, we observe that AlexNet on CPU exhibits a higher proportion of energy saving than ResNet152 and ViT-B/32 on GPU. This is because (1) GPUs are more energy-efficient during DNN inference than CPUs, leading to lower relative energy savings when switching to our method, and (2) as discussed in Section IV, the variance in inference time for ResNet152 and ViT-B/32 is relatively small, which limits the potential for our method to further reduce energy consumption. These results demonstrate that our proposed algorithms are particularly effective for DNNs with considerable variations in inference time.

Additionally, it can be seen from Figs. \ref{fig:AlexNet}(a), \ref{fig:ResNet152}(a), and \ref{fig:ViT-B/32}(a) that the energy consumption of mean-value optimization is lower than that of Algorithm 2, but mean-value optimization lacks probabilistic guarantees. It will be discussed in detail in the following.

\subsubsection{Impact of task deadlines}
Figs. \ref{fig:AlexNet}(b), \ref{fig:ResNet152}(b), and \ref{fig:ViT-B/32}(b) show total energy consumption at various task deadlines for a given risk level. It can be observed that for AlexNet, ResNet152, and ViT-B/32, the total energy consumption decreases monotonically as task deadlines increase. This is because, as the task deadline increases, mobile devices have more opportunities to select blocks with high computing power requirements and large inference time fluctuations for DNN partitioning and offloading these blocks to the MEC server for execution, thereby reducing local inference energy consumption. Additionally, the energy consumption of Algorithm 2 is lower than that of the worst-case policy at various task deadlines. For AlexNet, the energy consumption of Algorithm 2 decreases by $54.6\%$ when the deadline $D_n$ varies from 160 ms to 280 ms. For ResNet152, the energy consumption of Algorithm 2 decreases by $37.6\%$ when the deadline $D_n$ varies from 105 ms to 165 ms. For ViT-B/32, the energy consumption of Algorithm 2 decreases by $17.8\%$ when the deadline $D_n$ varies from 90 ms to 150 ms.

\subsubsection{Deadline violation probability}
Leveraging real-world data from Nvidia hardware platforms, we analyze the deadline violation probability of Algorithm 2 under various risk level settings. As illustrated in Fig. \ref{fig:AlexNet}(c), \ref{fig:ResNet152}(c) and \ref{fig:ViT-B/32}(c), we present the deadline violation probabilities at different risk levels achieved by Algorithm 2 and the other 2 benchmarks. As expected, the violation probabilities of the worst-case optimization for inference on AlexNet, ResNet152, and ViT-B/32 are all 0. It is because worst-case optimization is the most conservative method, but it also has the highest energy consumption, as shown in Figs. \ref{fig:AlexNet}(b), \ref{fig:ResNet152}(b), and \ref{fig:ViT-B/32}(b). As for the mean-value optimization, it leads to more than $40.5\%$, $49.5\%$, and $32.7\%$ violation probabilities for AlexNet, ResNet152, and ViT-B/32, respectively. This is due to the fact that the mean-value optimization approach fails to account for uncertainty in inference time. Although it achieves lower energy consumption compared to Algorithm 2, it is a primitive method that bypasses the uncertainty, without any probabilistic guarantees.

Then, we observe that the violation probability of Algorithm 2 is always lower than the risk level, which affirms the desired probabilistic guarantees and demonstrates the robustness of Algorithm 2 in handling uncertain DNN inference time. The gap between the risk level and the violation probability can be attributed to the fact that the actual inference time of DNNs does not invariably result in the maximum violation probability. This observation is consistent with our design in Section IV-A, where we approximate the variance of DNN inference time using the mean value in the CPU/GPU frequency scaling range. Although this approximation introduces some errors, it further reduces energy consumption compared with the maximum variance approximation while showing high robustness. It can be seen that as the risk level gradually increases, although the violation probability rises, it is still far below the risk level. As shown in Figs. \ref{fig:AlexNet}(a) and \ref{fig:ResNet152}(a), Algorithm 2 compared with worst-case optimization can achieve nearly $50.4\%$ and $9.4\%$ energy savings for AlexNet and ResNet152, respectively, when the actual violation probability is less than $0.5\%$ (i.e., $\varepsilon_n = 0.06$). Furthermore, Fig. \ref{fig:ViT-B/32}(a) shows that Algorithm 2 achieves about $6.4\%$ energy savings at an actual violation probability of $0.5\%$ (i.e., $\varepsilon_n = 0.03$).

\subsubsection{Performance under dynamic workload}
\begin{figure}[t]
  \captionsetup{font={small}}
  \centering
  \includegraphics[width=0.35\textwidth]{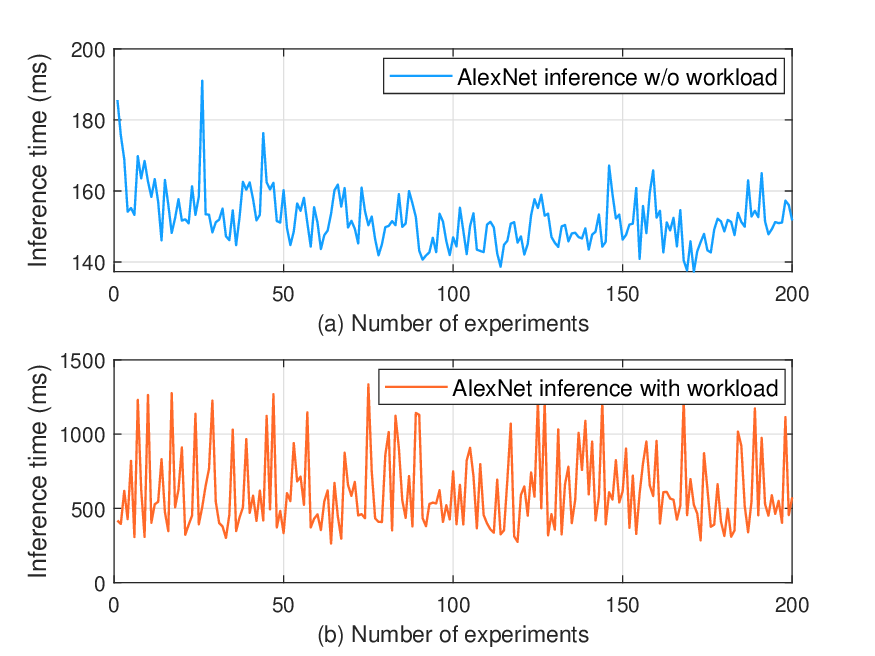}\\
  \caption{The impact of dynamic workload on inference time and its variation. (a) AlexNet executes local inference without any workload. (b) AlexNet executes local inference under a dynamic workload.}
  \label{fig:Workload}
  \vspace{-0.3cm}
\end{figure}

\begin{figure}[t]
  \captionsetup{font={small}}
  \centering
  \includegraphics[width=0.35\textwidth]{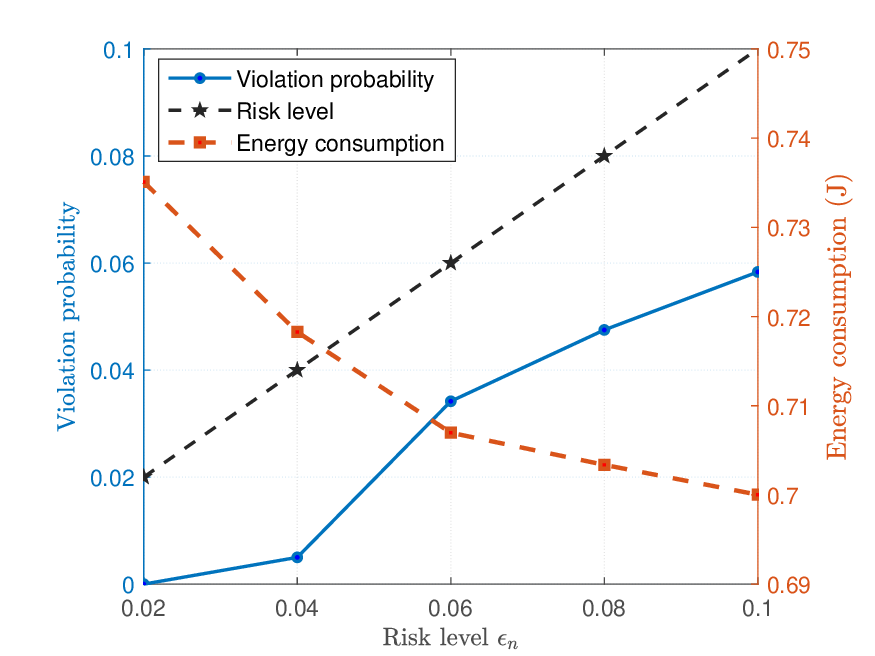}\\
  \caption{Deadline violation probability and energy consumption under different risk levels in a dynamic workload with $D_n=350$ ms, $B=10$ MHz, and $N=12$.}
  \label{fig:PerforWorkload}
  \vspace{-0.3cm}
\end{figure}

Fig. \ref{fig:Workload}(a) and Fig. \ref{fig:Workload}(b) show the changes in latency and variance of local inference executed by AlexNet before and after the application of a workload. In Fig. \ref{fig:Workload}(a), only AlexNet inference is executed on Jetson Xavier NX CPU, with no other workload. In Fig. \ref{fig:Workload}(b), while the AlexNet inference task is being executed, a trajectory planning task is also added to the CPU for parallel processing. It can be found that the mean inference time and the variation in inference time of AlexNet increase significantly when a workload is added.

Fig. \ref{fig:PerforWorkload} shows the performance of the proposed policy under the dynamic workload. It can be observed that as $\varepsilon_n$ increases, energy consumption gradually decreases. Moreover, even when the inference time fluctuates significantly, the actual deadline violation probability is lower than the risk level. This is because, through reasonable model partitioning, the system offloads DNN blocks that are heavily affected by the workload to the MEC server, thereby reducing the impact of high workload on the inference task.

\section{Conclusion}
In this paper, we investigated the problem of edge-device collaborative inference under uncertain inference time. Our experiments demonstrate that executing DNN inference tasks on high-performance GPUs can significantly enhance inference speed and reduce variations in inference time. This motivates us to develop an effective scheme for DNN model partitioning and resource allocation to achieve a balance among communication costs, computational requirements, and variations in inference time within edge intelligence systems. Therefore, we formulate the problem as an optimization problem that minimizes the total energy consumption of mobile devices while meeting task probabilistic deadlines. To solve this problem, we employ chance-constrained programming (CCP), which permits occasional violations of the target capacity threshold with a low probability, thereby reformulating the probabilistic constraint problem as a deterministic optimization problem. Then, the optimal solution of local CPU/GPU frequencies and uplink bandwidth allocation and a stationary point of DNN partitioning decisions are obtained using convex optimization and penalty convex-concave procedure (PCCP) techniques, respectively. We evaluate our proposed algorithm with real-world data and widely used DNN models. Extensive simulations demonstrate that, compared to worst-case optimization, our proposed algorithm achieves approximately $50.4\%$, $9.4\%$, and $6.4\%$ energy savings for AlexNet, ResNet152, and ViT-B/32, respectively, while maintaining an actual violation probability of not exceeding $0.5\%$.

Finally, we conclude the paper with several promising directions for future research. First, edge intelligence systems involve high-speed mobile scenarios, such as connected vehicles and drones. These mobile devices introduce uncertainties due to their dynamic nature. Consequently, our method can be extended to high-speed mobile scenarios and subject to joint design and optimization. Second, in multi-user edge intelligence systems, edge servers may experience computational resource competition when generating virtual machines (VMs) for each device, thereby introducing additional uncertainty-induced delays. Therefore, joint optimization and allocation of device-edge resources can further enhance system performance. Third, the proposed algorithm can be further extended to cross-continuous inference tasks, such as video streaming.  To achieve this, time-related information needs to be incorporated into the system model, and the impact of autocorrelation on inference time uncertainty and the probabilistic constraints in the CCP model should be analyzed.

\baselineskip=18pt
\bibliographystyle{IEEEtran}

\ifCLASSOPTIONcaptionsoff
  \newpage
\fi

\vspace{-8 mm}

\begin{IEEEbiography}[{\includegraphics[width=1in,height=1.25in,clip,keepaspectratio]{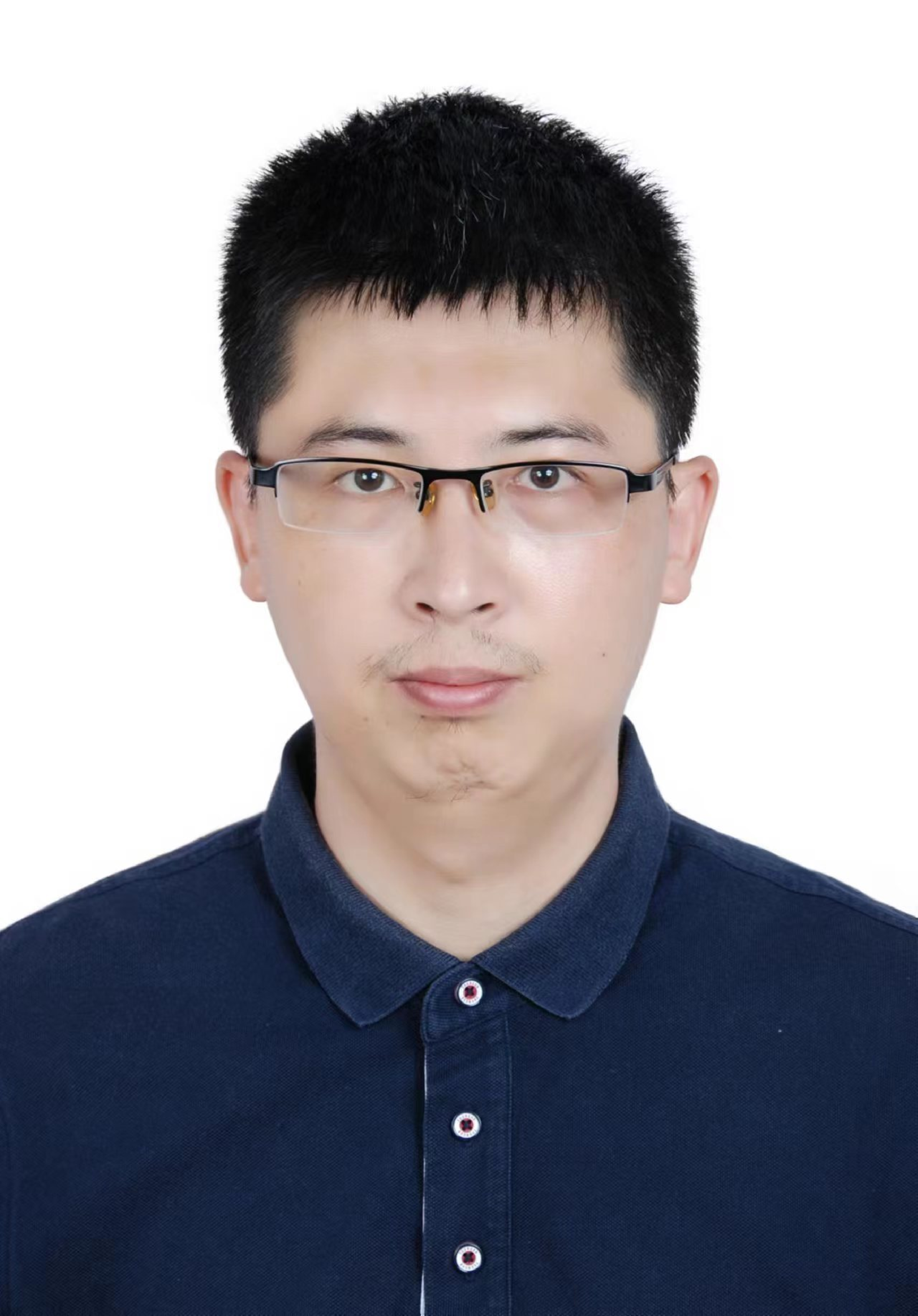}}]{Zhaojun Nan} (S'20-M'23) received the B.S. degree in automation from Jiamusi University, Jiamusi, China, in 2009, the M.S. degree in navigation, guidance and control from Harbin Engineering University, Harbin, China, in 2012, and the Ph.D. degree in information and communication engineering from Chongqing University, Chongqing, China, in 2022. From 2012 to 2017, he worked for Tianjin 712 Communication $\&$ Broadcasting Co.,Ltd., Tianjin, China. He is currently a Postdoctoral Researcher with the Network Integration for Ubiquitous Linkage and Broadband Laboratory, Department of Electronic Engineering, Tsinghua University, Beijing, China. His research interests include mobile edge computing, vehicular networks and autonomous driving, and green wireless communications. He has served as the TPC member for IEEE Globecom, ICC, VTC and WCNC.
\end{IEEEbiography}

\vspace{-8 mm}

\begin{IEEEbiography}[{\includegraphics[width=1in,height=1.25in,clip,keepaspectratio]{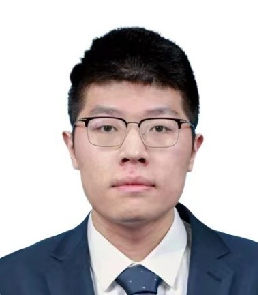}}]{Yunchu Han} (S'24) received the B.E. degree from the Department of Electronic Engineering, Tsinghua University in 2023. He is currently pursuing the Ph.D. degree with the Network Integration for Ubiquitous Linkage and Broadband Laboratory, Department of Electronic Engineering, Tsinghua University. His research interests include edge computing, edge intelligence, vehicular networks and green wireless communication.
\end{IEEEbiography}

\vspace{-8 mm}

\begin{IEEEbiography}[{\includegraphics[width=1in,height=1.25in,clip,keepaspectratio]{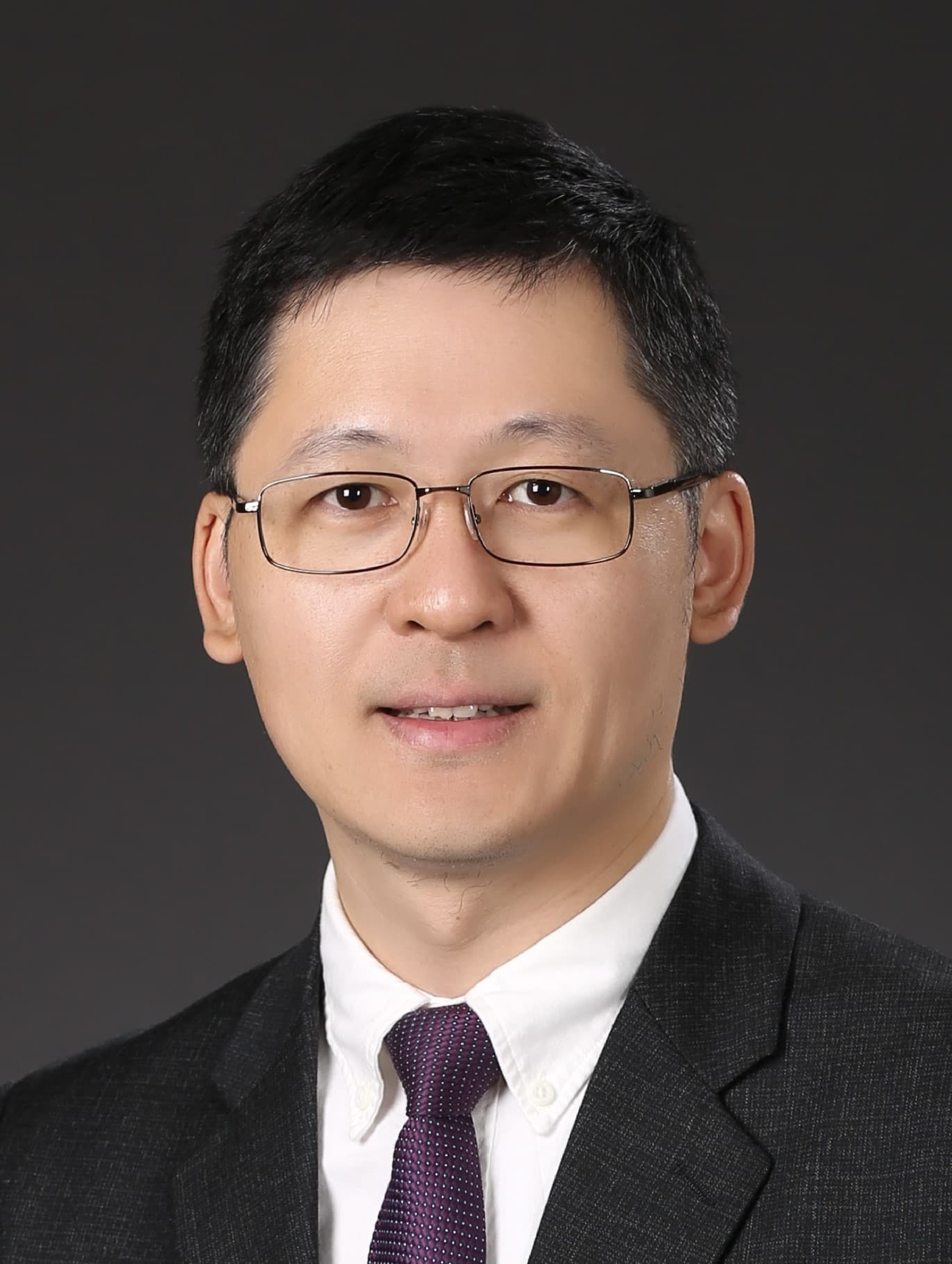}}]{Sheng Zhou} (S'06-M'12) received the B.E. and Ph.D. degrees in electronic engineering from Tsinghua University, Beijing, China, in 2005 and 2011, respectively. In 2010, he was a Visiting Student with the Wireless System Lab, Department of Electrical Engineering, Stanford University, Stanford, CA, USA. From 2014 to 2015, he was a Visiting Researcher with Central Research Lab, Hitachi Ltd., Tokyo, Japan. He is currently an Associate Professor with the Department of Electronic Engineering, Tsinghua University, Beijing, China. His research interests include cross-layer design for multiple antenna systems, mobile edge computing, vehicular networks, and green wireless communications. He was the recipient of the IEEE ComSoc Asia-Pacific Board Outstanding Young Researcher Award in 2017, and IEEE ComSoc Wireless Communications Technical Committee Outstanding Young Researcher Award in 2020.
\end{IEEEbiography}

\vspace{-8 mm}

\begin{IEEEbiography}[{\includegraphics[width=1in,height=1.25in,clip,keepaspectratio]{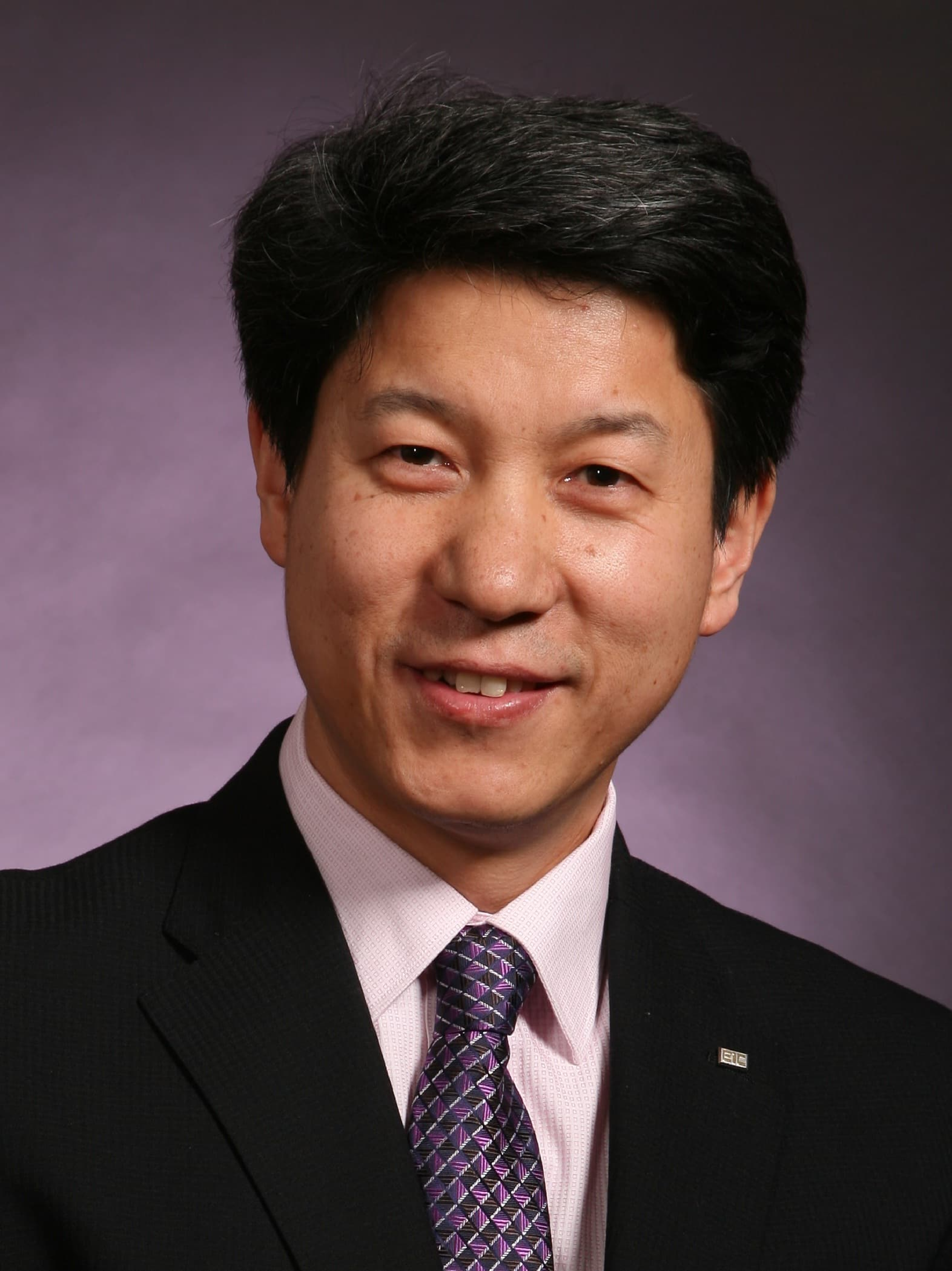}}]{Zhisheng Niu} (M'98-SM'99-F'12) graduated from Beijing Jiaotong University, China, in 1985, and got his M.E. and D.E. degrees from Toyohashi University of Technology, Japan, in 1989 and 1992, respectively.  From 1992 to 1994, he worked for Fujitsu Laboratories Ltd., Japan, and in 1994 joined with Tsinghua University, Beijing, China, where he is now a Professor at the Department of Electronic Engineering. His major research interests include queueing theory, traffic engineering, mobile Internet, radio resource management of wireless networks, and green communication and networks.

Dr. Niu has been serving IEEE Communications Society since 2000 as Chair of Beijing Chapter (2000-2008), Director of Asia-Pacific Board (2008-2009), Director for Conference Publications (2010-2011), Chair of Emerging Technologies Committee (2014-2015), Director for Online Contents (2018-2019) and the Editor-in-Chief of \textsc{IEEE Transactions on Green Communications and Networking} (2020-2022). He received the Best Paper Award of Asia-Pacific Board in 2013, Distinguished Technical Achievement Recognition Award of Green Communications and Computing Technical Committee in 2018, and Harold Sobol Award for Exemplary Service to Meetings \& Conferences in 2019, all from the IEEE Communications Society. He was selected as a distinguished lecturer of IEEE Communications Society (2012-2015) as well as IEEE Vehicular Technologies Society (2014-2018). He is a fellow of both IEEE and IEICE.
\end{IEEEbiography}

\end{document}